\documentclass[fleqn,usenatbib]{mnras}

\usepackage{newtxtext,newtxmath}

\usepackage[T1]{fontenc}

\DeclareRobustCommand{\VAN}[3]{#2}
\let\VANthebibliography\thebibliography
\def\thebibliography{\DeclareRobustCommand{\VAN}[3]{##3}\VANthebibliography}

\usepackage{graphicx}	
\usepackage{amsmath}	
\usepackage{amssymb}	

\usepackage{multirow} 
\usepackage{xcolor} 

\defcitealias{ws14}{WS14}
\defcitealias{lh11}{LH11}

\title[The Local Hole]{The Local Hole: a galaxy under-density covering 90\% of sky to $\approx200$ Mpc}

\author[J. H. W. Wong et al.]{
Jonathan H. W. Wong$^{1,2}$,\thanks{E-mail: jonathan.wong-6@postgrad.manchester.ac.uk}
T. Shanks$^1$,\thanks{E-mail: tom.shanks@durham.ac.uk}
N. Metcalfe$^1$
\&
J.R. Whitbourn$^1$
\\
$^1$ Centre for Extragalactic Astronomy, Department of Physics, Durham University, South Road, Durham, DH1 3LE, England.\\
$^2$ Jodrell Bank Centre for Astrophysics, Department of Physics and Astronomy, University of Manchester, Oxford Road, Manchester M13 9PL, UK.
}

\date{Accepted 2022 February 7. Received 2022 February 2; in original form 2021 July 18}

\pubyear{2022}

\begin{document}
\label{firstpage}
\pagerange{\pageref{firstpage}--\pageref{lastpage}}
\maketitle

\begin{abstract}

We investigate the `Local Hole', an anomalous under-density in the local
galaxy environment, by extending our previous  galaxy $K-$band
number-redshift and number-magnitude counts to $\approx 90$\% of the
sky. Our redshift samples are taken from the 2MASS Redshift Survey
(2MRS) and the 2M++ catalogues, limited to $K<11.5$. We find that both
surveys are in good agreement, showing an $\approx 21-22\%$
under-density at $z<0.075$ when compared to our homogeneous counts model
that assumes the same luminosity function and other parameters as in our
earlier papers. Using the Two Micron All Sky Survey (2MASS) for $n(K)$
galaxy counts, we measure an under-density relative to this model of $20
\pm 2 \%$ at $K<11.5$, which is consistent in both form and scale with
the observed $n(z)$ under-density. To examine further the accuracy of
the counts model, we compare its prediction for the fainter $n(K)$
counts of the Galaxy and Mass Assembly (GAMA) survey. We further compare
these data with a model assuming the parameters of a previous study
where little evidence for the Local Hole was found. At $13<K<16$ we find
a significantly better fit for our galaxy counts model, arguing for our 
higher luminosity function normalisation. Although our  implied
under-density of $\approx 20\%$ means local measurements of the Hubble
Constant have been over-estimated by $\approx3$\%, such a scale of
under-density is in tension with a global $\Lambda$CDM cosmology at an
$\approx3\sigma$ level.

\end{abstract}




\section{Introduction}

Distance scale measurements of the expansion rate of the Universe or
Hubble's Constant, $H_{0}$, have improved significantly over recent
years. For example, estimates of $H_{0}$ calculated by \citet{riess2016}
find a best fit value of $H_{0} = 73.24 \pm
1.74$\,km\,s$^{-1}$\,Mpc$^{-1}$, a quoted accuracy of 2.4\%. However,
this result is in serious tension with $H_{0}$ predictions made through
$\Lambda$CDM model fits to the Planck CMB Power Spectrum. This `early
Universe' measurement yields a value of $H_{0} = 67.4 \pm
0.5$\,km\,s$^{-1}$\,Mpc$^{-1}$ \citep[][]{planck2018}, which presents a
tension at the $3-4 \sigma$ level with measurements made using the local
distance scale \citep[see also][]{riess18b}.

These authors recognise the possibility that a source of the $\sim 9\%$ discrepancy between the $H_{0}$ measurements is unaccounted systematic uncertainties in one of, or both of the distance scale and early Universe approaches. However, an alternative proposal lies in studies of the galaxy distribution in the local Universe by \citet{shanks90}, \citet{Metcalfe91}, \citet{metcalfe01}, \citet{frith03} \& \citet{busswell04}, who find evidence for an under-density or `Local Hole' stretching to $150-200\,h^{-1}$\,Mpc in the local galaxy environment. 

Notably, \citet{ws14} (hereafter \citetalias{ws14}) suggest that the tension in $H_{0}$ measurements may arise from the outflow effects of the Local Hole. They find a detected under-density of $\approx 15\pm3\%$ in number-magnitude counts $n(m)$ and redshift distributions $n(z)$, measured relative to a homogeneous model over a $\sim 9,000$ square degree area covering the NGC and SGC. This under-density is most prominent at $K<12.5$ and leads to an $\sim 2-3\%$ increase in $H_{0}$ which alleviates the tension to a $5\%$ level. Further, \citet{shanks19} suggested that Gaia DR2 parallaxes might not have finally confirmed the Galactic Cepheid distance scale as claimed by \cite{riess18b} and could at least superficially, help  reduce the overall tension to $<1 \sigma$.

Moreover, the existence of the Local Hole has been detected in wider
cluster distributions, with \citet{bohringer15}, \citet{collins16} and \citet{bohringer20}
finding underdensities of $\sim 30\%$ in the  X-Ray cluster redshift
distributions of the REFLEX II and CLASSIX surveys respectively. These
results are in strong agreement with the galaxy counts of
\citetalias{ws14}, and suggest that the observed $H_0$ within the
under-density would be inflated by $5.5^{+2.1}_{-2.8}\%$.

Contrastingly, \citet{riess18a} critique the assumption of isotropy and spherical symmetry assumed in the modelling of the  Local Hole, highlighting that the \citetalias{ws14} dataset covers only $20\%$ of the sky, yet measurements drawn from this subset are projected globally to draw conclusions on the entire local environment. These authors further suggest that such an all-sky local under-density would  then be incompatible with the expected cosmic variance of mass density fluctuations in the $\Lambda$CDM model at the $\approx6 \sigma$ level. In addition, \citet{kenworthy19}  failed to find dynamical evidence in the form of infall velocities for the Local Hole in their Pantheon supernova catalogue. 

Further, through  analyses of the galaxy distribution in the 2M++
Catalogue, \citet{jasche19}, following \citet{lh11} (hereafter
\citetalias{lh11}) find that local structure can be accommodated within
a standard concordance model, with no support for an under-density on
the scale suggested by \citetalias{ws14}. However,  \citet{shanks19b}
(see also \citealt{ws16}) question the choice of the Luminosity Function
(LF) parameters used by \citet{jasche19} and \citetalias{lh11}.

In this work we will examine two aspects of the above arguments against
the Local Hole. First, to address the premise that the conclusions of
\citetalias{ws14} cover too small a sky area to support a roughly
isotropic under-density around our position, we will extend the analysis
of \citetalias{ws14} and measure $K-$ band $n(m)$ and $n(z)$ galaxy
counts over $\approx90$\% of the sky to a limiting Galactic latitude
$|b|\ga 5^\circ$.

Second, we will compare the $n(z)$ and $n(K)$ model predictions of
\citetalias{ws14} with \citet{lh11},  hereafter \citetalias{lh11}. These
predictions will be compared at both the bright 2MASS limit and at the
fainter $K-$band limit of the GAMA survey to try and understand the
reasons for the different conclusions of \citetalias{ws14} and \citetalias{lh11} on the existence of the `Local Hole'.

\section{Data}
\label{sec:data}

\subsection{Photometric Surveys}
\label{subsec:photometric_surveys}

We now detail properties of the photometric surveys used to provide
$n(m)$ counts, alongside calibration techniques and star-galaxy
separation methods that we apply to ensure consistency between the
photometric data and model fit. Following \citetalias{ws14}, we choose
to work in the Vega system throughout. Thus for the GAMA
survey, we apply a $K-$band conversion from the $AB$ system according to
the  relation determined by \citet{Driver2016}:

\begin{equation}
K_s(Vega) = K_s(AB)-1.839
\end{equation}

\subsubsection{2MASS}
\label{sub_subsec:2mass}

The Two Micron All Sky Survey, 2MASS \citep{skrutskie06} is a
near-infrared photometric survey achieving a 99.998\% coverage of the
celestial sphere. In this work we will take $K$-band $n(m)$ counts from
the 2MASS Extended Souce Catalogue (2MASS$\_$xsc), which is found to be
$\sim97.5\%$ complete \citep{mcintosh06}, with galaxies thought to
account for $\approx 97\%$ of sources.

For the galaxy $n(m)$ results, we choose to work in Galactic
coordinates, and present counts from down to a limiting Galactic
latitude $|b|>5^{\circ}$ except for the Galactic longitude range,
$330<l<30^\circ$ where our limit will be $|b|>10^{\circ}$. This is the
same 37063 deg$^2$ area of sky used by \citetalias{lh11}. These cuts are
motivated by the increasing density of Galactic stars at lower latitudes
and close to the Galactic Centre.

Following \citetalias{ws14}, sources are first selected according to the
quality tags `\textit{cc\_flg}=0' or `\textit{cc\_flg}=Z'.  We will work
with a corrected form of the 2MASS$\_$xsc extrapolated surface
brightness magnitude, `$K\_m\_ext$', quoted in the Vega system.  The
conversion we use is detailed in \citetalias{ws14} Appendix A1, and
utilises the $K-$band photometry of \citet{loveday2000}. For sources in
the range $10<K<13.5$ we take a corrected form of the magnitude,
$K\_\text{Best}$, defined as:

\begin{equation}
K\_\text{Best} = 0.952 \times (K\_m\_ext + 0.5625)
\end{equation}

\noindent The effect of converting to the $K\_\text{Best}$ system is to
slightly steepen the observed counts at the fainter end. However, the
effect of the conversion is small and its inclusion does not alter the
conclusions we draw.

To remove stellar sources in 2MASS we exploit here the availability of
the Gaia EDR3 astrometric catalogue \citep{Gaia2016,Gaia2021} and simply
require that a source detected in Gaia EDR3 is not classed as pointlike
as defined  by eq \ref{eq:gaia} of Section
\ref{sub_subsec:star_separation}. But when compared to the star-galaxy
separation technique used by WS14, little difference to the galaxy
$n(K)$ and $n(z)$ is seen.

Finally, 2MASS galaxy $K_s$ magnitudes are corrected throughout for
Galactic absorption using the $E(B-V)$ extinction  values determined by
\citet{Schlafly2011} and $A_{K_s}=0.382E(B-V)$. The coefficient
here corresponds to the relation $A_V=3.1E(B-V)$ for the $V$-band.

\subsubsection{GAMA}
\label{sub_subsec:gama}

The Galaxy And Mass Assembly, GAMA  survey \citep{driver09} provides a
multi-wavelength catalogue covering the near- and mid-infrared,
comprising $\approx 300,000$ galaxies over an area of $\approx180$
deg$^2$. The survey offers deeper $K$ counts which are not accessible in
the 2MASS sample, so we will use GAMA to compare the ability of the
\citetalias{ws14}- and \citetalias{lh11}-normalised models to fit faint
$K-$band $n(m)$ counts. Measurements will be taken from the GAMA DR3
release \citep{baldry18} using the Kron magnitude
`\textit{MAG\_AUTO\_K}', initially given in the AB system. We will
target the combined count of the 3 equatorial regions G09, G12 and G15,
each covering 59.98 square degrees with an estimated galaxy completeness
of $\approx 98.5\%$ \citep{baldry10}. We shall take the GAMA sample to
be photometrically complete to $K<15.5$ but only complete to $K<15$ for
their redshift survey since a visual inspection of the $K$ counts of
galaxies with redshifts indicate that only the G09 and G12 redshift
surveys reach this limit. For star-galaxy separation we shall first use 
the $g-i:J-K$ galaxy colour-based method recommended for GAMA by
\citet{baldry10} (see also \citealt{Jarvis13}) before applying the  Gaia
criteria of Section \ref{sub_subsec:star_separation} to this subset to
reject any remaining stars.

\subsubsection{VICS82}
\label{sub_subsec:vics82}

VISTA-CFHT Stripe 82, VICS82 \citep{Geach2017}, is a survey in the
near-infrared over $J$ and $K_s$ bands, covering $\approx150$ deg$^2$ of
the SDSS Stripe82 equatorial field. The survey provides deep coverage to
$K<20$. Sources are detected and presented measuring a total magnitude
‘$MAG-AUTO$’ quoted in the AB system. The image extraction gives a
star-galaxy separation flag, `$Class~Star$’, with extended and
point-like sources distributed at 0 and 1 respectively. Whereas
\citet{Geach2017} defined pointlike sources at $Class~Star> 0.95$, we
shall define extended objects using a more conservative cut at
$Class~Star<0.9$. We then use the Gaia method of  Section
\ref{sub_subsec:star_separation} to remove any remaining pointlike
objects. In terms of $K$ magnitude calibration, we start from the same
VICS82 $K\_mag\_auto$ system as \citet{Geach2017}  who note that there
is zero offset to 2MASS total $K\_20$ magnitudes (see their Fig. 4).
However, in Appendix \ref{appendix_vics82} we find that between $12.0<K\_m\_ext<13.5$
the offset $K\_m\_ext-K\_VICS82=0.04\pm0.004$ mag and this is the offset we  use 
for these VICS82 data in this work. As with GAMA, we then use the
deep $K-$ band counts of VICS82 to test how well the  \citetalias{ws14}
model predicts faint galaxy counts beyond the 2MASS $K<13$ limit.

\subsubsection{Star-Galaxy Separation using Gaia}
\label{sub_subsec:star_separation}

The Gaia Survey \citep{gaia18} provides an all-sky photometry and
astrometry catalogue for over 1 billion sources in the $G-$band, and is
taken as essentially complete for stars between $G=12$ and $G=17$. The
filter used to determine pointlike objects makes use of the total flux
density `$G$' and astrometric noise parameter `$A$', which is a measure
of the extra noise per observation that can account for the scatter of
residuals \citep{lindegren18}. Explicitly, through the technique of
\citet{krolewski20}, pointlike sources are then classified as:

\begin{equation}
\label{eq:gaia}
\text{pointlike}(G,A)=\left\{
	\begin{array}{ll}
		\log_{10}A<0.5  & \mbox{if } G<19.25 \\
		\log_{10}A<0.5 + \frac{5}{16}(G-19.25) & \mbox{otherwise}
	\end{array}
\right.
\end{equation}
This separation technique is applied, sometimes in combination with other
techniques, to the raw photometric datasets taken from 2MASS, GAMA and
VICS82 used to analyse the wide-sky and faint-end $n(m)$ counts.

\subsection{Redshift Surveys}
\label{subsec:redshift_surveys}

We now present characteristics of the redshift surveys used to measure
the $n(z)$ galaxy distribution, and the techniques we apply to ensure
the data remain consistent with those of \citetalias{ws14}. 
 
To achieve close to all-sky measurement, we similarly take the observed
$n(z)$ survey distribution to the same \citetalias{lh11} $(l,b)$ limits
discussed  in Section \ref{sub_subsec:2mass}, and work with redshifts
reduced to the Local Group barycentre (see Eq. 10 of \citetalias{ws14}).
While \citetalias{ws14} use the SDSS and 6dFGRS surveys to measure
separate distributions in the northern- and southern-galactic
hemispheres respectively, we will access a larger sky area using the
wide-sky redshift surveys based on  the photometric 2MASS catalogue.

\subsubsection{2MRS} 
\label{sub_subsec:2MRS}

The 2MASS Redshift Survey, 2MRS \citep[][]{huchra12} is a spectroscopic
survey of $\sim45,000$ galaxies covering 91$\%$ of the sky built from a
selected sample of the 2MASS photometric catalogue limited to $K<11.75$.
The 2MRS Survey is reported to be 97.6$\%$ complete excluding the
galactic region $|b|<5^{\circ}$, and provides a coverage to a depth
$z\sim0.08$.

To remain consistent with the $n(m)$ distributions, we work with a
$K-$band limited 2MRS sample, achieved by matching the 2MRS data with
the star-separated 2MASS Extended Source Catalogue. To minimise
completeness anomalies, we take a conservative cut at $K<11.5$ to
measure the $n(z)$ distribution. In Table~\ref{tab:numbers} we provide
summary statistics of the $n(z)$ dataset achieved by the matching
procedure, alongside the corresponding 2MASS $n(m)$ count.

\subsubsection{2M++} 
\label{sub_subsec:2M++}

The 2M++ Catalogue \citep[][]{lh11} is a spectroscopic survey of
$\sim70,000$ galaxies comprised of redshift data from 2MRS, 6dFGRS and
SDSS. The 6dFGRS/SDSS and 2MRS data are given to $|b|>10^{\circ}$ and
$|b|>5^{\circ}$ respectively, except in the region
$-30^{\circ}<l<+30^{\circ}$ where 2MRS is limited to $|b|>10^{\circ}$.

The 2M++ Catalogue applies masks to this field to associate particular
regions to each survey, weighting by completeness and magnitude limits.
Overall, this creates a set of galaxies covering an all-sky area of
$37,080$ deg$^{2}$ which is thought to be $\sim 90\%$ complete to $K
\leq 12.5$. To compare with counts from 2MRS we will measure the
redshift distribution to a depth $K<11.5$, with the summary statistics
presented in Table~\ref{tab:numbers}.

\begin{table}
	\centering
	\caption{ Summary statistics of the $n(m)$ and $n(z)$ datasets we use for analysis of the Local Hole over the wide-sky area ($|b|>5^{\circ}$ except for  $|b|>10^{\circ}$ at $330^\circ<l<30^\circ$).}
	\label{tab:numbers}

	\begin{tabular}{ccccc}
	\hline
	\hline
	\multirow{2}{*}{Survey}	& 	Wide-Sky Area				& 	\multirow{2}{*}{Mag. Limit}	& 	$n(m)$ 		& 	\multirow{2}{*}{$n(z)$} 		\\
							&	(sq. deg.)					&								&	(2MASS)						&				\\
	\hline
	\hline
	2MRS 					&	\multirow{2}{*}{$37,063$}	&	\multirow{2}{*}{$K<11.5$}	&	\multirow{2}{*}{$41,771$}	&	$38,730$		\\
	2M++						&								&								&								&	$34,310$		\\
	\hline
	2MRS 					&	\multirow{2}{*}{$37,063$}	&	\multirow{2}{*}{$K<11.75$}	&	\multirow{2}{*}{$59,997$}	&	$43,295$		\\
	2M++						&								&								&								&	$44,152$		\\	
	\hline
	\hline
	\end{tabular}
\end{table}

\subsubsection{Spectroscopic Incompleteness}
\label{sub_subsec:spectroscopic_incompleteness}

For a given $n(z)$ sample taken from 2MRS and 2M++, we correct the data
using an incompleteness factor. The observed $n(z)$ distribution from
the survey is multiplied by the ratio of the total number of photometric
to spectroscopic galaxies within the same target area and magnitude
limit. Here, the photometric count is taken from the 2MASS Extended
Source Catalogue and the correction ensures that the total number of
galaxies considered in the redshift distribution $n(z)$ is the same as
in the magnitude count $n(m)$. A breakdown of the completeness of each
survey as a function of magnitude is presented in Appendix \ref{appendix_spectra}.

\subsection{Field-field errors}
\label{subsec:error_calculation}

The field-field error, $\sigma$, in the galaxy 2-D sky or 3-D volume
density in each photometric or spectroscopic bin is simply calculated by
sampling the galaxy densities in $n$  sub-fields within the
wide-sky area and calculating their standard error. For $n$
sub-fields each with galaxy density, $\rho_i$, the standard error
$\sigma$ on the mean galaxy density,  $\bar{\rho}$, in each magnitude or
redshift bin is therefore,

\begin{equation}
\sigma^{2}=\frac{1}{n(n-1)}\sum_{i=1}^{n}\left(\rho_i-\bar{\rho}\right)^{2}
\label{eq:error}
\end{equation}

\noindent So, for the 2MASS wide-sky survey, we divide its area into 20
subfields each covering 1570 deg$^2$ over the majority of the sky, but
in the offset strip for $330^\circ<l<30^\circ$, there are 4 additional
subfields of equal area 1420 deg$^2$ that have slightly different
boundaries. The 10\% smaller boundaries for 4 out of 24 sub-fields is
assumed to leave eq \ref{eq:error} a good approximation to the true
field-field  error estimate. In Section~\ref{sec:all_sky_nm} we detail
the Galactic coordinate boundaries of each sub-field in a Mollweide
projection, and consider the individual galaxy densities in each of
these $n=24$ sub-fields to visualise the extent on the sky of the Local
Hole.

\section{Modelling}
\label{subsec:modelling}

To examine the redshift and magnitude distribution of galaxies, we
measure their differential number counts per square degree on the sky as
a function of magnitude, $m$,  and redshift, $z$,  over a bin size
$\Delta m = 0.5$ and $\Delta z = 0.002$ respectively. The observed
counts are then compared to the \citetalias{ws14} theoretical
predictions that assumed a model based on the sum of contributions from
the type-dependent LFs of \citet{metcalfe01}. The LF  parameters
$\phi^{*}, \alpha, M^{*}$, which represent the characteristic density,
slope and characteristic magnitude respectively, are presented for each
galaxy type in Table~\ref{tab:params}.

The apparent  magnitude of galaxies is further dependent on their
spectral energy distribution and evolution, modelled through $k(z)$ and
$e(z)$ corrections respectively. Thus, we calculate the apparent
magnitude $m$ by including these in the distance modulus for the $[m,z]$
relation:

\begin{equation}
m = M + 5\log_{10}(D_L(z)) + 25  + k(z) + e(z) 
\end{equation}

\noindent where $D_L(z)$ represents the luminosity distance at redshift,
$z$. In this work, the $k$ and $e$ corrections are adopted from
\citetalias{ws14} who adopt Bruzual \& Charlot (2003) stellar synthesis
models. We note that the K band is less affected by $k$ and $e$
corrections than in bluer bands because of the  older stars that
dominate in the near-IR.

\begin{figure*}
	\includegraphics[width=17cm]{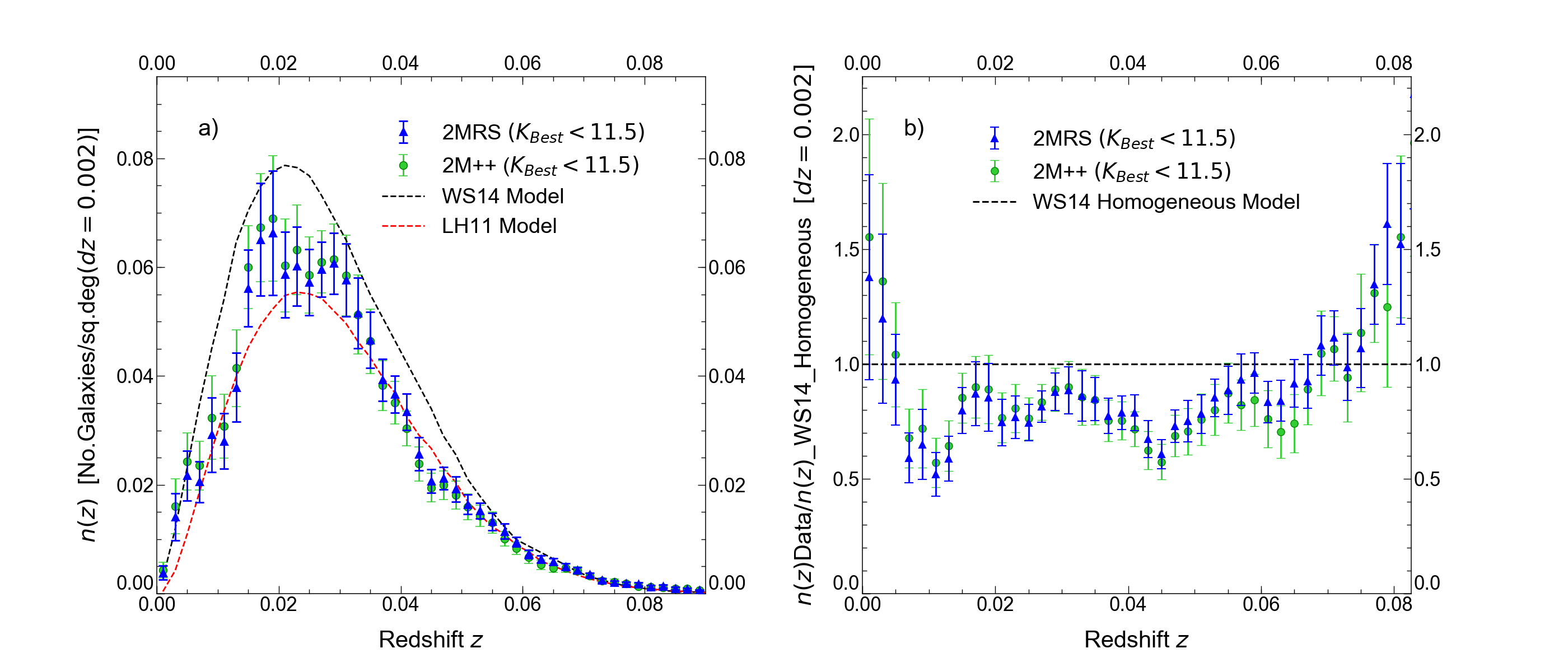}
	\caption{The observed $n(z)$ distributions of the 2MRS and 2M++
	Catalogues measured to the wide-sky area $|b|\ga5^\circ$,
	and consistently limited to $K<11.5$ where: \textbf{(a)} Counts are
	fit alongside the WS14 homogeneous model and LH11-normalised model
	over a bin size $\Delta z=0.002$. \textbf{(b)} The $n(z)$ counts are
	normalised to the WS14 model to demonstrate observed under- and
	overdensities across the distribution.}
    \label{fig:nz}
\end{figure*}

In addition to the basic homogeneous prediction, we consider the
\citetalias{ws14} inhomogeneous model in which the normalisation
$\phi^{*}$ is described as a function of redshift. We trace the radial
density profile shown in each redshift bin of the observed  $n(z)$ count
(see Fig.~\ref{fig:nz} (b)), and apply this correction to the $n(m)$
model prediction according to

\begin{equation}
\phi^{*}(z)=
\begin{cases}
\frac{n(z)_{\text{obs}}}{n(z)_{\text{global}}}\phi^{*}_{\text{global}}	&	z \leq z _{\text{global}} \\
\phi^{*}_{\text{global}}													&	z > z_{\text{global}}
\end{cases}
\end{equation}

\noindent where the $n(z)_{\text{obs}}$ are the observed distributions
from our chosen redshift surveys, $\phi^{*}_{\text{global}}$ describes
the standard homogeneous normalisation as detailed in
Table~\ref{tab:params}, and $z_{\text{global}}$ is the scale at which
the inhomogeneous model transitions to the homogeneous galaxy density. 

In such a way we can model the effect of large-scale structure in the
number-magnitude prediction, which we use as a check for consistency in
measurements of any under-density between the observed $n(m)$ and $n(z)$
counts. In this work we test the effect of two transition values
$z_{\text{global}}=0.06$ and $z_{\text{global}}=0.07$.

\begin{table}
	\centering
	\caption{The luminosity function parameters defined at zero redshift as
     a function of galaxy-type, used as the homogeneous model by WS14 and
     adopted in this work. The absolute magnitudes are `total' $K-$band
     magnitudes, corresponding to our $K\_\text{Best}$ system. Here, the Hubble
     parameter $H_0=100\,h\,$km\,s$^{-1}$\,Mpc$^{-1}$.}
	\label{tab:params}
	\begin{tabular}{cccc}
	\hline
	\hline
	Type	& $\phi^{*}(h^{3}$\,Mpc$^{-3})$ & $\alpha$ & $M^*_K+5log_{10}(h)$ \\
	\hline
	\hline
	E/S0	&	$7.42\times10^{-3}$ &	$-0.7$		&	$-23.42$ \\
	Sab		&	$3.70\times10^{-3}$ &	$-0.7$		&	$-23.28$ \\
	Sbc		&	$4.96\times10^{-3}$ &	$-1.1$		&	$-23.33$ \\
	Scd		&	$2.18\times10^{-3}$ &	$-1.5$		&	$-22.84$ \\
	Sdm		&	$1.09\times10^{-3}$ &	$-1.5$		&	$-22.21$ \\
	\hline
	\hline
	\end{tabular}
\end{table}

\section{Galaxy redshift distribution}
\label{sec:all_sky_nz}

The observed $n(z)$ distribution measured in the 2MRS and 2M++
catalogues over the wide-sky area to $|b|\ga5^\circ$ is shown in
Fig.~\ref{fig:nz}(a). The data are limited to $K<11.5$ and compared to
the $n(z)$ predictions of the homogeneous \citetalias{ws14} LF
model,\footnote{We note that convolving the
\citetalias{ws14} model $n(z)$ with a Gaussian of width
$\sigma_z=0.001$ to represent the combined effect of redshift errors and
peculiar velocities of $\pm300$km s$^{-1}$ shows no discernible
difference.}  with a corresponding plot of the observed $n(z)$ divided
by the model shown in Fig.~\ref{fig:nz} (b). Counts have been corrected
with the spectroscopic incompleteness factor described in
Section~\ref{sub_subsec:spectroscopic_incompleteness}, and a description
of the completeness of each sample as a function of magnitude is given
in Appendix \ref{appendix_spectra}. Errors have been calculated using
the field-field method incorporating the uncertainty in each observed
redshift bin combined with the uncertainty in the incompleteness.

Subject to the limiting magnitude $K<11.5$, each survey shows a
distribution where the majority of the observed $n(z)$ data fall below
the predicted count of the \citetalias{ws14} homogeneous model. The
observed distributions fail to converge to the model until $z>0.06$ and
below this range the data exhibit a characteristic under-density that is
consistent with $n(z)$ counts over the NGC and SGC presented in
\citetalias{ws14}.

To analyse the scale of under-density in our measurements, we consider
the `total' density contrast, calculated by evaluating the difference
between the sum of the observed count and predicted count, normalised to
the sum of the predicted count.  Here we take the sum over $n(z)$ bins
from $z=0$ to the upper limits of $z=0.05$ and $z=0.075$. Calculations
of the density contrast in our wide-sky 2MRS and 2M++ distributions
within these bounds are presented in Table~\ref{tab:nz_all_sky}.

\begin{table}

\begin{center}

\caption{The measured density contrasts between the WS14 LF
model and $n(z)$ counts of 2MRS and 2M++ over the $\sim 37,000$ sq. deg.
wide-sky area. The samples are taken to a limiting magnitude $K<11.5$
and detail the scale of under- and overdensities to the specified ranges
$z<0.05$ and $z<0.075$.}
\label{tab:nz_all_sky}

\begin{tabular}{ccc}
\hline
\hline
Sample Limit & Survey & Density Contrast $(\%)$ \\
\hline
\hline
\multirow{2}{*}{$z<0.05$}	& 2MRS & $-23\pm2$ \\
						 	& 2M++ & $-21\pm3$ \\	
\hline
\multirow{2}{*}{$z<0.075$}	& 2MRS & $-22\pm2$ \\
						 	& 2M++ & $-21\pm2$ \\
\hline
\hline						 	

\end{tabular}

\end{center}

\end{table}

The measured density contrast of each survey at $z<0.075$ are in
excellent agreement and indicate that the wide-sky $n(z)$ counts are
$\sim 21-23\%$ underdense relative to the model. At both limits, the
2MRS dataset produces a marginally greater under-density than 2M++,
however, the two values remain consistent to within $1 \sigma$ and
demonstrate a continuous under-density in the $n(z)$ distribution. 

\begin{figure*}
	\includegraphics[height=8.5cm]{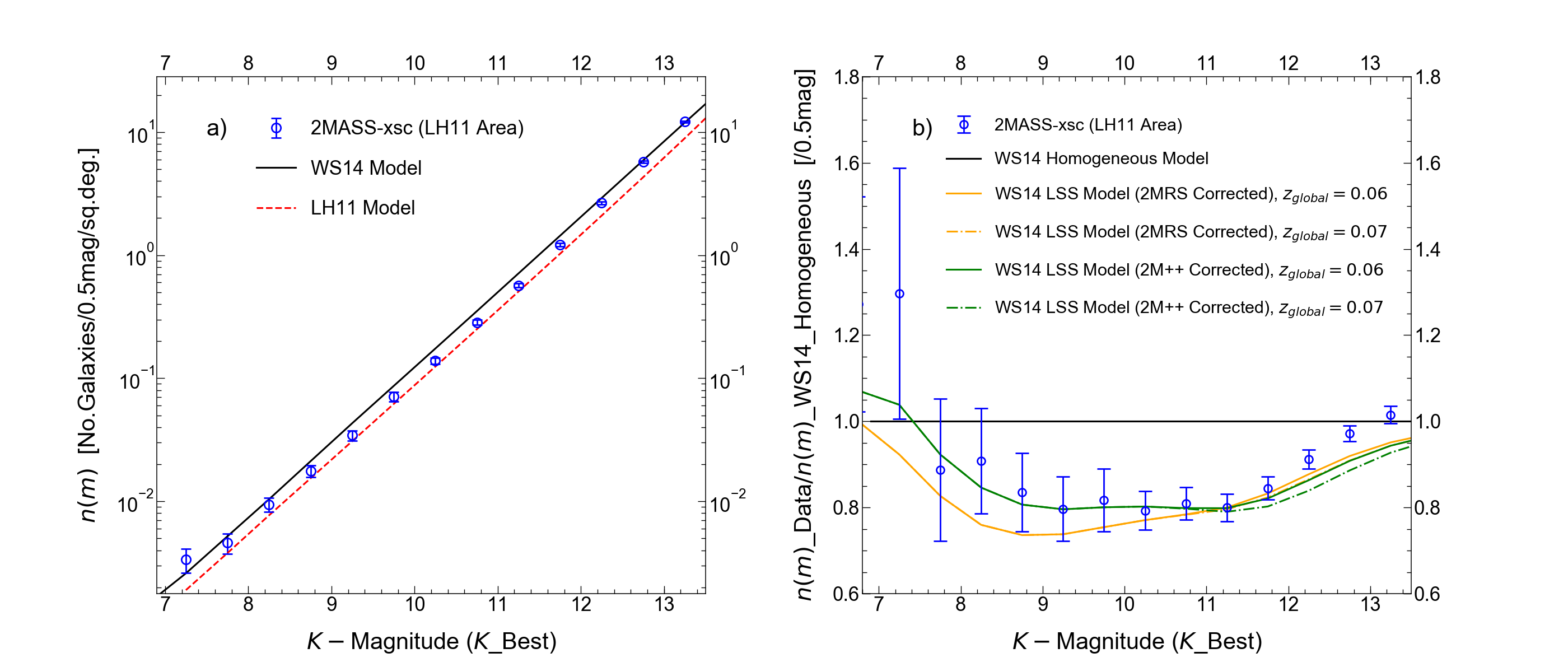}
	\caption{The observed $K-$band $n(m)$ counts of the 2MASS Extended
	Source Catalogue taken over the wide-sky region to
	$|b|\ga5^\circ$, where: \textbf{(a)} The observed counts
	are compared to the WS14 and LH11 homogeneous models. \textbf{(b)}
	The observed counts  divided by the WS14 homogeneous model are
	compared  to the inhomogeneous, variable $\phi^{*}(z)$, versions of
	the WS14 models based on the 2MRS and 2M++ $n(z)$'s, and similarly
	divided by the homogeneous WS14 model. The transition to the
	homogeneous case for both of these inhomogeneous LSS models is investigated 
	for both $z_{\text{global}}=0.06$ and $z_{\text{global}}=0.07$. }
    \label{fig:nm}
\end{figure*}

We note that in our approach we have applied a single incompleteness
factor to correct each bin in the observed $n(z)$ distribution equally
while a more detailed examination could incorporate a
magnitude-dependent factor. This technique was implemented in
\citetalias{ws14}, where the completeness factor was introduced into the
LF $n(z)$ model such that each bin conserved the galaxy number. However,
the change to the $n(z)$ sample as a result of this method was less than
$1\%$ and so we have not implemented this more detailed correction here.

We shall return in  Section \ref{sec:lh11_2mass}  to discuss
the reasons for the difference in the $n(z)$  model prediction of \citet{lh11},
also shown in Fig. \ref{fig:nz}(a). 
 
\begin{figure*}
	\includegraphics[width=17cm]{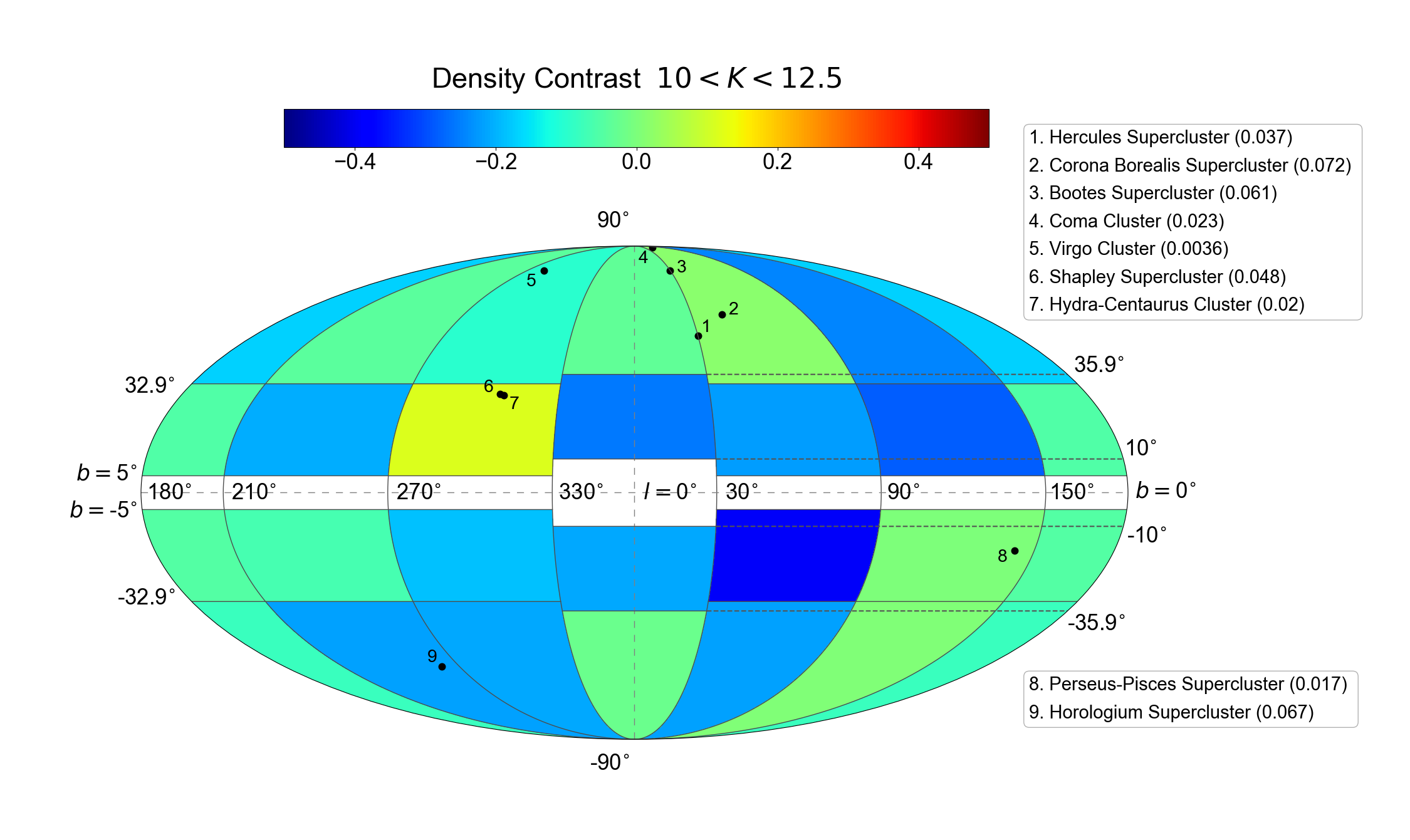}
	\caption{A Mollweide contour plot detailing the galactic coordinate
	positions of each sub-field we have used to calculate the
	field-field errors in our wide-sky $n(m)$ and $n(z)$ distributions.
	In each region we have evaluated the 2MASS $n(m)$ density contrast,
	measured at $10<K<12.5$, and plotted local galaxy structures to
	investigate the regional densities. The legend describes the key for
	each galaxy structure, and their corresponding redshift is given in
	brackets.}
    \label{fig:mollweide}
\end{figure*}

\section{Galaxy number magnitude counts}
\label{sec:all_sky_nm}

\subsection{2MASS $n(m)$ counts}
\label{subsec:2MASS_nm}

We now consider the 2MASS number-magnitude counts, and examine the
extent the \citetalias{ws14} homogeneous model can self-consistently
replicate an $n(m)$ under-density that is of the same profile and at a
similar depth as that suggested by the galaxy redshift distributions of
2MRS and 2M++.

The observed $K-$band $n(m)$ count of the 2MASS Extended Source
Catalogue to the wide-sky limit of $|b|\ga5^\circ$, is presented in
Fig.~\ref{fig:nm}(a). Similar to the $n(z)$ comparison in Fig.
\ref{fig:nz}, these counts appear low compared to the homogeneous model
of \citetalias{ws14}, here at $K<12$.

To examine whether the $n(m)$ counts are consistent with the form of the
under-density shown in the $n(z)$ measurements, we also predict  this
$n(m)$ based on the LSS-corrected $\phi^*(z)$ normalisation (see Section
\ref{subsec:modelling}). We  first show the observed $n(m)$ count
divided by the homogeneous \citetalias{ws14} model in
Fig~\ref{fig:nm}(b). Then we use the $n(z)_{\text{obs}}$ derived from
each of the 2MRS and 2M++ $n(z)$ distributions in Fig.~\ref{fig:nz} (b),
both similarly divided by the \citetalias{ws14} homogeneous model. The
orange and green lines represent the 2MRS and 2M++ -corrected models
respectively.

At $K<12.5$, the wide-sky $n(m)$ distribution shows a significant
under-density relative to the homogeneous prediction, only reaching 
consistency with the model at $K\approx13$. Moreover, we find that the
$\phi^{*}(z)$ models describing the observed $n(z)$ inhomogeneities in
each of 2MRS and 2M++ give a significantly more accurate fit to the
2MASS $n(m)$ count. This indicates that the profile of the under-density
in the galaxy redshift distributions, measured relative to the
\citetalias{ws14} homogeneous prediction, is consistent with the
observed $n(m)$ counts.

To explicitly evaluate the 2MASS $n(m)$ under-density, we give
calculations of the density contrast in Table~\ref{tab:nm_all_sky}. To
mitigate the uncertainty at the bright end and remain in line with
measurements given by \citetalias{ws14}, we take a fixed lower
bound, $K>10$, and vary the upper magnitude bound.

\begin{table}
\begin{center}
\caption{Measurements of the density contrast in the 2MASS wide-sky
$n(m)$ counts relative to the WS14 model, taken to various $K-$ limits
to examine the extent of underdensities in the distribution. Errors are field-field
based on 24 sub-fields.}
\label{tab:nm_all_sky}
\begin{tabular}{ccc}
\hline
\hline
Sky Region & Sample Limit & Density Contrast ($\%$) \\
\hline
\hline
$|b|\ga5^\circ$ & $10<K<11.5$ & $-20 \pm 2$ \\
$|b|\ga5^\circ$ & $10<K<12.5$ & $-13 \pm 1$ \\
$|b|\ga5^\circ$ & $10<K<13.5$ & $-3 \pm 1$ \\
\hline
\hline
\end{tabular}
\end{center}
\end{table}

The measurements of the total density contrast in the wide-sky $n(m)$
count in Table \ref{tab:nm_all_sky} demonstrate a significant scale of
under-density at $10<K<11.5$ that becomes less pronounced approaching
$K\approx13.5$. Notably, at $K<11.5$ we measure an under-density of
$20\pm2\%$, which is consistent with the $\approx 21-22\%$ under-density
shown in the 2MRS and 2M++ $n(z)$ counts. Additionally, for $K<12.5$ we
find a wide-sky under-density of $13\pm1\%$, which is in good agreement
with the $15\pm3\%$ under-density calculated in the three
\citetalias{ws14} fields over the same magnitude range. The field-field
errors suggest strongly significant detections of a 13-21\% underdensity
over the wide-sky area. This is in agreement with \citetalias{ws14}, who
found an $\approx15$\%  underdensity from their sample covering a
$\approx4\times$ smaller area over the NGC and SGC. In addition, we note
the effect of the magnitude calibration to the Loveday system. Excluding
the correction lowers the observed count at the faint end by
$\approx10\%$, confirming the conclusion of \citetalias{ws14} that an
under-density is seen independent of applying the Loveday magnitude
correction. Finally, we again shall return in  Section
\ref{sec:lh11_2mass} to discuss why the  $n(m)$  model prediction of
\citetalias{lh11} also shown in Fig. \ref{fig:nm}(a) are so much lower than
that of \citetalias{ws14}.

\subsection{Sub-field $n(m)$ Density Contrast Measurements}
\label{subsec:Mollweide_Plot}

To further assess the sky extent of the Local Hole, we next consider the
properties of the wide-sky sub-fields from which we derive the
field-field errors and evaluate the 2MASS $n(m)$ density contrast in
each sub-field region.

Fig.~\ref{fig:mollweide} shows the density contrast between the 2MASS
$n(m)$ counts and the \citetalias{ws14} homogeneous model in each
sub-field area that is also used to evaluate the wide-sky $n(m)$ and $n(z)$
field-field errors. The average density contrast  in each field is plotted
colour-coded on a Mollweide projection, which also details the geometric
boundaries of each region in Galactic coordinates. 

To probe the under-density, we choose to take the sum over the range
$10<K<12.5$  to remain consistent with the limits considered in
\citetalias{ws14}. In addition, to examine the properties of individual
regions we plot the local galaxy clusters and superclusters highlighted
in \citetalias{lh11} using positional data from \citet{abell89},
\citet{einasto97} and \citet{ebeling98}, and provide their redshift as
quoted by \citet{huchra12} in the 2MRS Catalogue.

From the lack of yellow-red colours in  Fig.~\ref{fig:mollweide} it is
clear that underdensities dominate  the local Large Scale Structure
across the sky. Now, there are several fields which demonstrate an
$n(m)$ count that marginally exceeds the \citetalias{ws14} prediction
and we find that such (light green) regions tend to host well known
local galaxy clusters. The 4 out of 24 areas that show an over-density
are those that contain clusters 2,3,4 - Corona Borealis+Bootes+Coma; 6,7
- Shapley+Hydra-Centaurus; 8 - Perseus-Pisces, using the numbering system
from Fig. \ref{fig:mollweide}. The influence of the structures in these 4
areas is still not enough to dominate the Local Hole overall $13\pm1$\%
under-density in the wide-sky area in Fig.~\ref{fig:mollweide}.

 \begin{figure*}
	\includegraphics[width=\columnwidth]{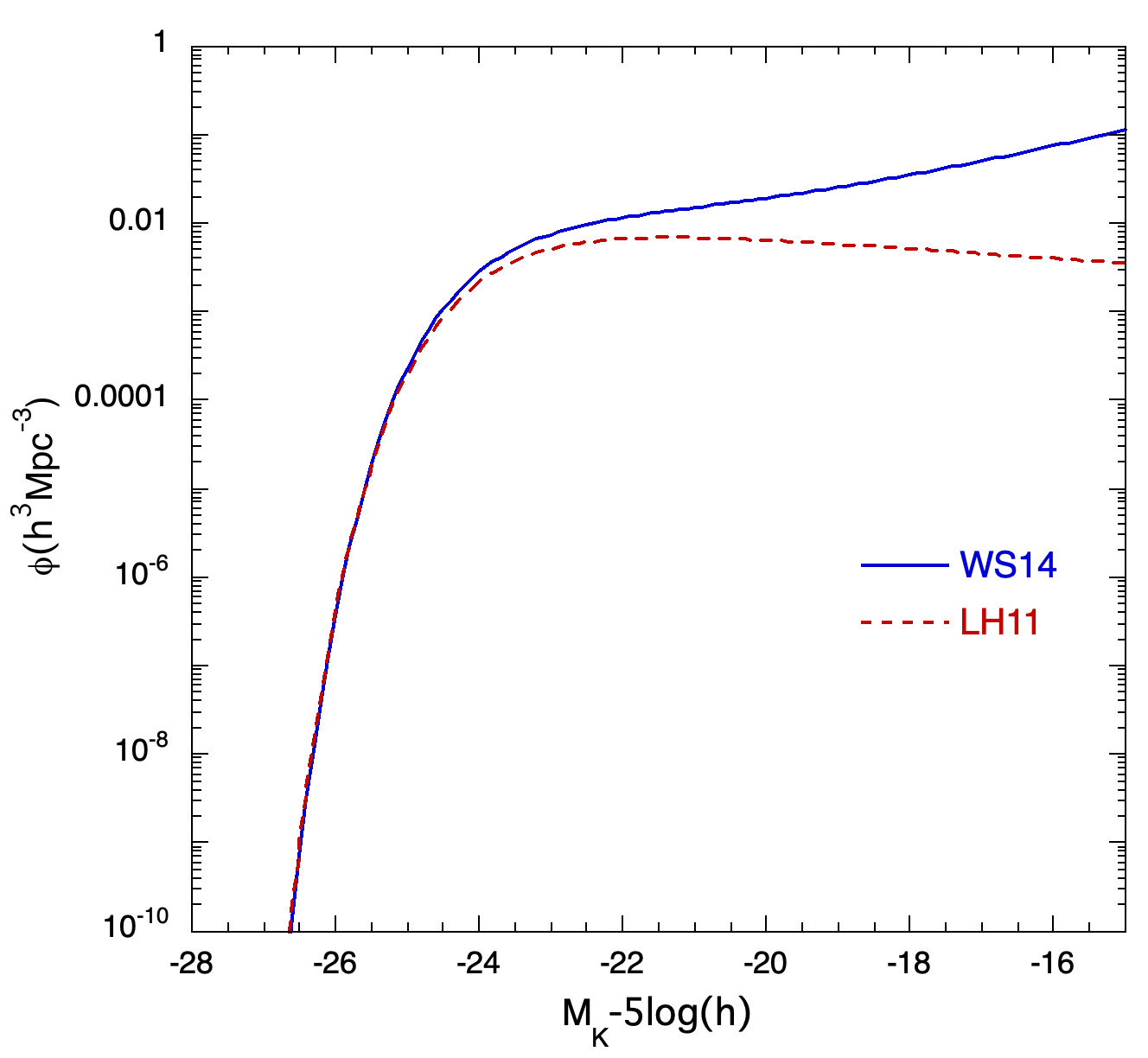}	
	\includegraphics[width=8.5cm]{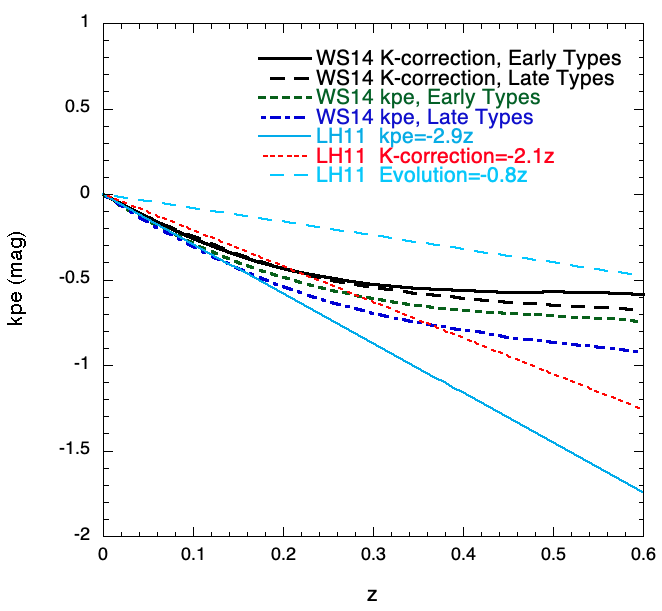}
	\caption{(a) The  galaxy $K$ luminosity function of WS14 as used here
	compared to that of LH11. (b) The $k$ and $k+e$ corrections of WS14
	compared to those of LH11, for the $K$-band.}
    \label{fig:LF_params}
\end{figure*}

We conclude that the observed $n(m)$ and $n(z)$ galaxy counts taken to
$|b|\ga5^\circ$ in 2MASS, 2MRS and 2M++, show a consistent overall
under-density measured relative to the \citetalias{ws14} model that
covers $\approx90$\% of the sky. At a limiting depth of $K=11.5$ the $n(m)$ counts
show an under-density of $20\pm2$\% and this scale is replicated in form
in the $K-$limited $n(z)$ distributions at $z<0.075$ which show an
under-density of $\sim21-22$\%.

\section{Comparison of LF and other model parameters}
\label{sec:normalisation_testing}

The above arguments for the Local Hole under-density depend on the
accuracy of our model LF and to a lesser extent our $k+e$ parameters
that are the basis of our $n(z)$ and $n(m)$ models. We note that
\citet{ws16} made several different estimates of the galaxy LF in the
$K$ band from the $K<12.5$ 6dF and SDSS redshift surveys including
parametric and non-parametric `cluster-free' estimators and found good
agreement with the form of the LF used by \citetalias{ws14} and in this
work. The `cluster-free' methods are required since they ensure that at
least the form of the LF is independent of the local large-scale
structure and mitigates the presence of voids as well as clusters. The
non-parametric estimators also allowed independent estimates of the
local galaxy density profiles to be made and showed that the results of
\citetalias{ws14} were robust in terms of the choice of LF model. The
\citetalias{ws14} LF normalisation was also tested using various methods
as described in Section 2.3.1 of \citet{ws16}.

We now turn to a comparison between the \citetalias{ws14} galaxy count
predictions with those made by \citet{lh11} who failed to find an
under-density in the 2M++ $n(z)$ data. To examine the counts produced by
their model  we assume the LF parameters given in their Table 2, where
in  the Local Group frame with $750<v<20000$ km s$^{-1}$, they find
$\alpha=-0.86$; $M^*=-23.24+5{\rm log}_{10}(h)$;
$\phi^*=1.13\times10^{-2} h^3 {\rm Mpc}^{-3}$, independent of galaxy
type. Note that we brighten the LH11 $M^*$ by 0.19 mag to
$M^*=-23.43+5{\rm log}_{10}(h)$ in our version of their model to account
for the 0.19 mag difference between $K\_m\_ext$ magnitudes used here and
the 2MASS $K_{Kron}(=K\_20)$ magnitudes used by \citetalias{lh11} (see
Appendix \ref{appendix_2mass}). In Fig. \ref{fig:LF_params} (a) we
compare their $z=0$ LF with our LF summed over our five galaxy types.
Importantly, \citetalias{lh11} note that their fitted LFs show a
distinctly flatter faint slope ($\alpha>-1$) than other low redshift LF
estimates (see their Fig 7a) that generally look more similar to the
steeper \citetalias{ws14} LF (see also \citealt{ws16}).  However,
Fig.~\ref{fig:LF_params} (a) shows that the form of both LF's is similar
in the range around $M^*$ that dominates in magnitude limited galaxy
samples, apart from their normalisation, with the  \citetalias{lh11} LF
appearing $\approx40$\% lower than that of \citetalias{ws14}. We shall argue
that this low normalisation is crucial in  the failure of
\citetalias{lh11} to find the `Local Hole'.

Next, we compare the $k+e$ - redshift models of \citetalias{lh11} and
\citetalias{ws14} in Fig. \ref{fig:LF_params}b. Two $k+e(z)$ models are
shown for \citetalias{ws14} representing their early-type  model applied
to E/S0/Sab and their late type model applied to Sbc/Scd/Sdm. These
models come from \cite{bc03} with parameters as described by
\cite{metcalfe06} At $z=0.1$ these models give respectively
$\Delta_K=-0.28$ and $\Delta_K=-0.31$. We also show just the $k(z)$ for
early and late types in Fig.\ref{fig:LF_params}.  At $z=0.1$ these
$k(z)$ models give respectively $\Delta_K=-0.26$ and $\Delta_K=-0.25$,
implying little evolution in the $e(z)$ model for the early types and
0.06 mag for the late types.

We note that \citetalias{lh11} apply their $k+e$ corrections to the data
whereas we apply them to the model. So reversing their sign on their
$k(z)$ and $e(z)$ terms, the correction we add to our $K$ magnitudes in
our count model is

\begin{equation}
\Delta_K(z)=k(z)-e(z).
\end{equation}

\citetalias{lh11} give $k(z)=-2.1z$ and $e(z)=0.8z$ giving our additive correction as

\begin{equation}
\Delta_K(z)=k(z)-e(z)=-2.1z-0.8z=-2.9z
\end{equation}

\noindent as representing the \citetalias{lh11} $k$- and evolutionary
corrections, giving  $\Delta_K=-0.29$mag at $z=0.1$. Their second model
includes an additional galaxy $(1+z)^4$ surface brightness dimming
correction so in magnitudes is 

\begin{equation}
\Delta_K(z)=0.16(10\log_{10}(1+z))+1.16(k(z)-e(z))
\end{equation}

\noindent i.e.

\begin{equation}
\Delta_K(z)=1.6\log_{10}(1+z)-3.4z
\end{equation}

\noindent and so $\Delta_K=-0.27$mag at $z=0.1$.


Since we are using total $K$ magnitudes, the effect of cosmological
dimming of surface brightness is included in our measured magnitudes. So
in any comparison of the \citetalias{lh11} model with our $K$ band data,
only the $k+e$ terms are used in the model.  So at $z=0.1$ our $k+e$
term is $\Delta_K\approx-0.29$mag, the same as the  $\Delta_K=-0.29$mag
of \citetalias{lh11}. Similarly at $z=0.3$ which is	effectively  our
largest redshift of interest at $K<15.5$, $z=0.3$,
$\Delta_K\approx-0.60$ to -0.69 mag for the \citetalias{ws14} $k+e$
model compared to $\Delta_K=-0.87$mag for \citetalias{lh11}.

\subsection{Lavaux \& Hudson $n(m)$ and $n(z)$ comparisons to $K=11.5$}
\label{sec:lh11_2mass}

In Figs.~\ref{fig:nz}(a) and ~\ref{fig:nm}(a) we now compare the
\citetalias{lh11} model predictions to those of \citetalias{ws14} for
the 2MRS and 2M++ $n(z)$ and 2MASS $n(K)$ distributions. Most notably,
we find that the \citetalias{lh11} model produces theoretical $n(K)$ and
$n(z)$ counts that are significantly lower than the \citetalias{ws14}
counterparts and, if anything, slightly  {\it under}-predict the observed
wide-sky counts particularly near the peak of the $n(z)$ in
Fig.~\ref{fig:nz}(a). The \citetalias{lh11} $n(K)$ model is offset by
$\approx40$\% from the \citetalias{ws14} $n(K)$ prediction. We also note
that the $n(z)$ distribution predicted  by \citetalias{lh11} when
compared to  the  2M++ $n(z)$, limited at $K=11.5/12.5$ mag, shows
excellent  agreement (see \citetalias{lh11} Fig. 5). However, in
Fig.~\ref{fig:nm}(a), beyond $K>12.5$, the \citetalias{lh11} $n(K)$ model
diverges away from the 2MASS data. In contrast, the \citetalias{ws14}
model was  found to generate a consistency between the wide-sky $n(K)$
and $n(z)$ distributions and imply a similar under-density of
$\approx20$\% at $K<11.5$.  Due to the consistency of the slope in each
model at both the bright and faint end of $n(K)$ counts, it is likely
that the difference between the \citetalias{lh11} and the
\citetalias{ws14} models is caused by the different effective
normalisation in $\phi^*$ seen around the break in the LF in
Fig.~\ref{fig:LF_params}.

We further note that when we try to reproduce Fig. 5 of
\citetalias{lh11}, by combining $n(z)$ model predictions using their LF
model parameters for their combined $K<11.5$ and $K<12.5$ 2M++ samples
covering 13069 and 24011 deg$^2$ respectively, we find that we
reasonably reproduce the form and normalisation of their predicted
$n(z)$ to a few percent accuracy. So why the fit of the
\citetalias{lh11} model is poorer than in our Fig. \ref{fig:nz} (a) than
in their Fig. 5 remains unknown. Nevertheless, we accept that their
model fits our Fig. \ref{fig:nz} (a) $n(z)$ better than the model of
\citetalias{ws14}.

\subsection{Lavaux \& Hudson $n(m)$ comparison at $K<16$}

\begin{figure}
    \includegraphics[width=\columnwidth]{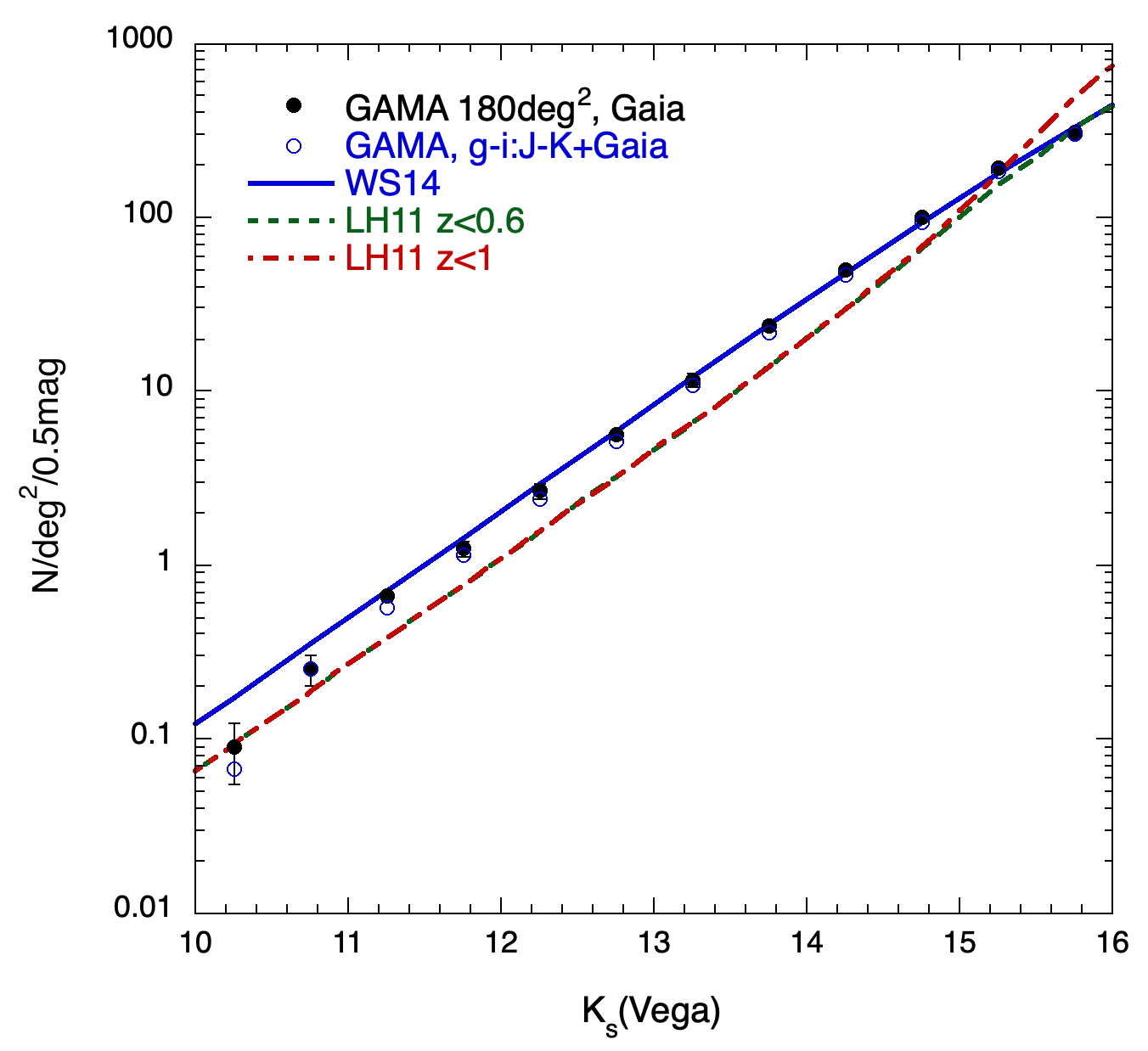}     
    \caption{The WS14 and  LH11 count models  compared to the GAMA
    survey observed $n(K)$ counts averaged over 3 fields. Solid circles
    are the GAMA counts with the Gaia star-galaxy separation and open
    circles are with the Gaia separation  applied after star-galaxy
    separating by colour \protect\citep{baldry10}. Two versions of the
    LH11 model are shown with redshift cuts at $z<0.6$ and $z<1.0$ to
    prevent the model diverging due to an unphysical high redshift tail.
    Field-field errors based on the 3 GAMA fields are shown.}
    \label{fig:gama}
\end{figure}

To examine the ability of the \citetalias{lh11} model simultaneously to
predict the galaxy $n(K)$ at bright and faint magnitudes, we now compare
the \citetalias{lh11} and \citetalias{ws14} models to the fainter 
$n(K)$ counts of the GAMA survey, shown in Fig.~\ref{fig:gama}. We
calculate errors using field-field errors as described in Section
\ref{subsec:error_calculation}.

To compare the count models, we again assume the \citetalias{lh11} LF
parameters from their Table 2, $\alpha=-0.86$; $M^*=-23.24+5{\rm
log}_{10}(h)$ (corrected brighter by 0.19 mag into our system);
$\phi^*=1.13\times10^{-2} h^3 {\rm Mpc}^{-3}$.  We also  assume the $k+e$ term
of $\Delta_K=-2.9z$ used by \citetalias{lh11}, one cut at $z<0.6$ and
one cut at $z<1$ as shown by the dashed and dotted lines in Fig.
\ref{fig:gama}.

The two \citetalias{lh11} predictions reasonably fit the bright data at
$K<11$ but lie below the observed GAMA data out to $K\approx15$, then
agreeing with these data at $K\approx15.5$. In the case of the version
cut at $z<1.0$, the model then rises above the GAMA counts. The model
cut at $z<0.6$ remains in better agreement with these data. But without
the redshift cuts we find that the $\Delta K=-2.9z$ k+e term used by
\citetalias{lh11} would vastly overpredict the observed galaxy count not just at
$K>15.5$mag but at brighter magnitudes too. This is the usual problem
with an evolutionary explanation of  the steep count slope at $K<12$, in
that models that fits that slope then invariably overpredict the
slope at fainter magnitudes. For an evolutionary model to fit, a strong
evolution, either in galaxy density or luminosity (as in the
\citetalias{lh11} + \citetalias{ws14} models used here) is needed out to
$z<0.1$ and then something quite close to a no-evolution  model is
required at $0.1<z<1$ in the $K$ band. This is similar to what was found
in the $b_J$-band where strong luminosity evolution is at least more
plausible. In $K$ the evolution is less affected by increasing numbers
of young blue stars with redshift and so the evolutionary explanation is
even less attractive.

The conclusion that the steep $K$ counts are caused by local large-scale
structure rather than evolution is strongly supported by the form of the
$n(z)$ seen in Fig. \ref{fig:nz} where the pattern of underdensities is
quite irregular as expected if dominated by galaxy clustering rather
than the smoothly increasing count with $z$ expected from evolution. We
have also shown that following the detailed changes in $n(z)$ with
redshift to model  $\phi^*(z)$ gives a consistent fit to the steep
$n(m)$ distribution at $K<12$. We conclude that unless a galaxy
evolution model appears that has the required quick cut-off at
$z\approx0.1$ required simultaneously in the $K$ and $b_J$ counts then
the simplest explanation of the steep $n(K)$ slope at bright magnitudes
is the large scale structure we have termed the `Local Hole`.

\begin{figure}
  	\includegraphics[width=\columnwidth]{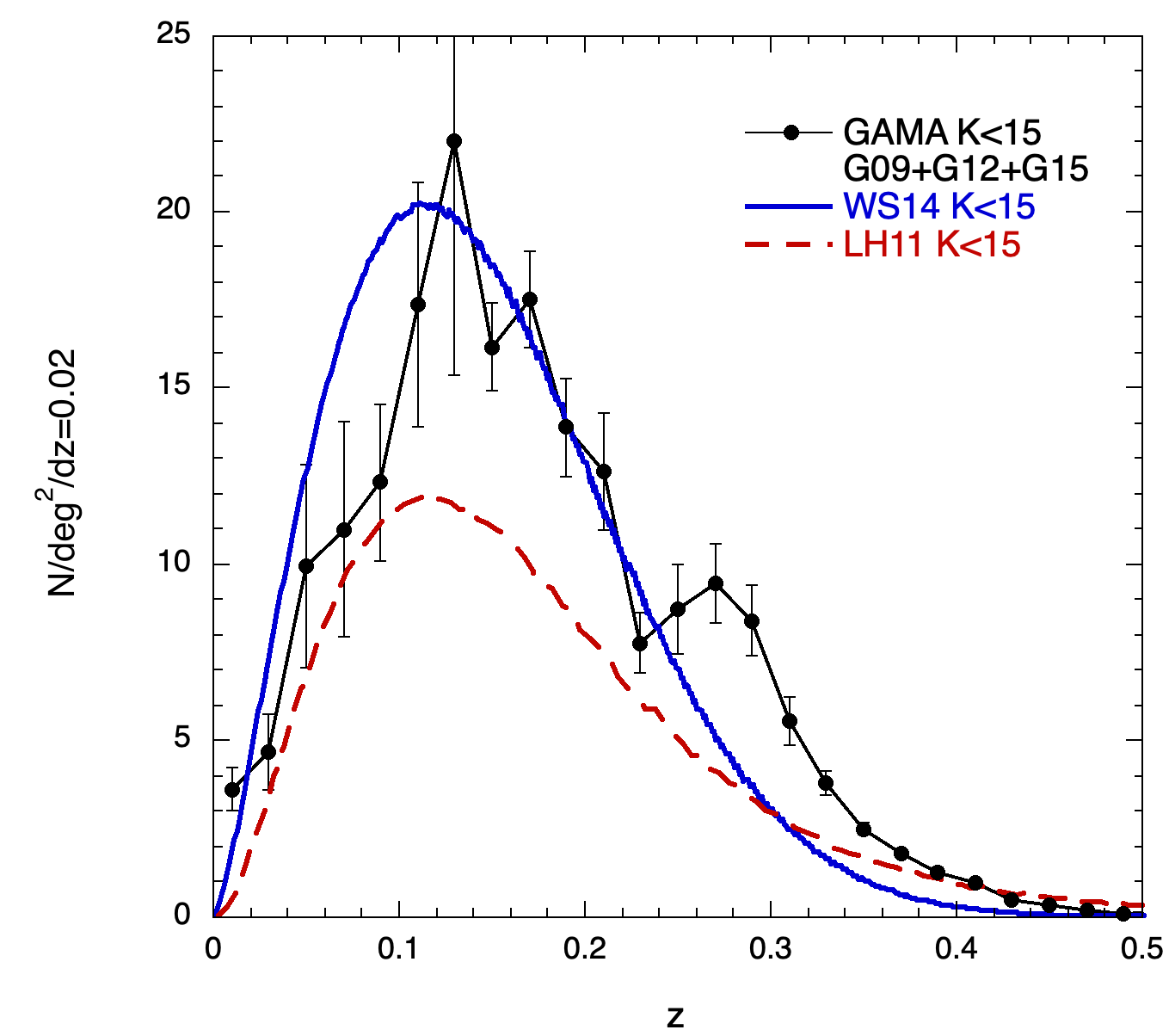}	
 	\caption{Galaxy $n(z)$ for GAMA survey limited at $10<K<15$ and the
    predictions of the WS14 and LH11  models. We chose the $K<15$ limit here
    because this appears to be the effective limit for the $K$ band
    spectroscopic survey in G09 and G12, although G15 may be complete to a
    0.5mag fainter limit. We note that there is a `bump' in the GAMA $n(z)$ at $z\approx0.25$
    that appears to have its origin mostly in the G09 and G15 fields with less contribution from  
    G12. G09 and G15  are the two most widely separated fields of the three,
    arguing that this feature is a statistical fluctuation, if not caused by some 
    $z$ survey target selection issue. 
    }
    \label{fig:gama_nz}
\end{figure}

These conclusions are confirmed by the GAMA $n(z)$ in the range
$10<K<15$, averaged over the G09, G12 and G15 fields and compared to the
\citetalias{ws14} + \citetalias{lh11} models in Fig. \ref{fig:gama_nz}.
Similar results are seen to those for the GAMA $n(K)$ in Fig.
\ref{fig:gama} with the \citetalias{ws14} model better fitting these
data than the \citetalias{lh11} model that again significantly
underestimates the observed $n(z)$. Some hint of an under-density is
seen out to  $z\approx0.12$ in the \citetalias{ws14} model
comparison with the observed data but the area covered is only $180$
deg$^2$ so the statistical errors are much larger than for the brighter
$K<11.5$ or $K<12.5$ `wide-sky' redshift survey
samples.\footnote{We note that at the suggestion of a
referee, we investigated the 2MASS Photometric Redshift Survey 
(2MPZ, \citealt{Bilicki2014}) $n(z)$ over the wide sky area used in Fig.\ref{fig:nz} to
$K<13.7$, finding evidence that this underdensity may extend to
$z\approx0.15$. But since this result could be affected by as yet unknown
systematics in the 2MPZ photometric redshifts, we have left this analysis for
future work.}

\begin{figure}
\includegraphics[width=\columnwidth]{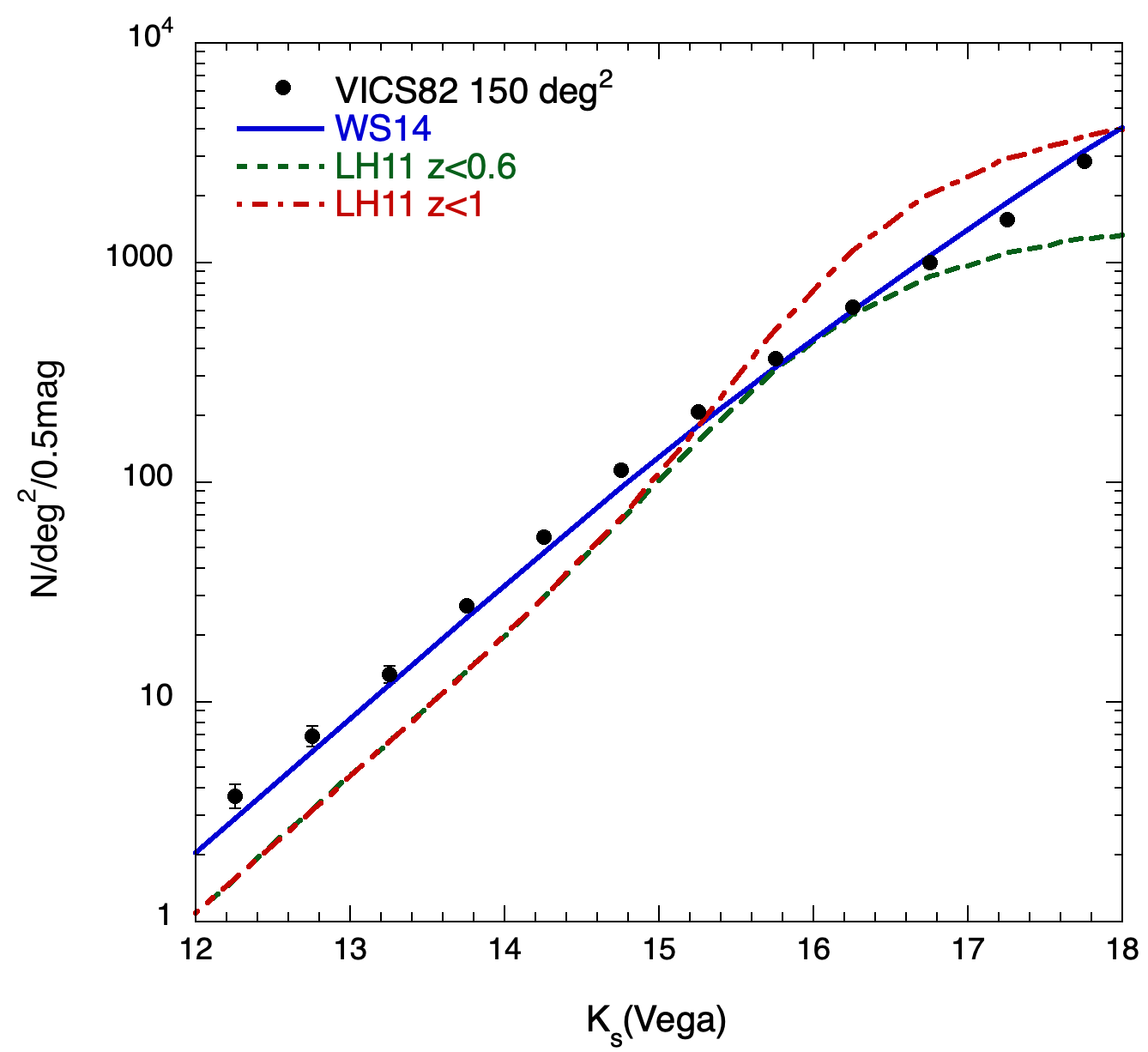}
\caption{The WS14 and  LH11 count models  compared to the VICS82 survey
\protect\citep{Geach2017} observed $n(K)$ counts averaged over
$\approx150$ deg$^2$ to $K<18$. Results are based on star-galaxy
separation $Class\_Star<0.9$ with further removal of Gaia pointlike
objects as defined by eq \ref{eq:gaia}. Field-field errors based on 2
sub-fields of area 69 and 81 deg$^2$ are shown.  The \citetalias{lh11}
models again have  redshift cuts at $z<0.6$ and $z<1$ to prevent 
divergence due to an unphysical high redshift tail.}
\label{fig:vics82} 
\end{figure}

\subsection{VICS82 $K$ count model comparison to $K=18$}

To assess further the LF normalisation uncertainties, we present  in
Fig. \ref{fig:vics82} the  $n(K)$ galaxy counts in the range $12<K<18$
over the $\approx150$deg$^2$ area of the VICS82 survey
\citep{Geach2017}. Here, the faint $K=18$ limit is 2 mag fainter than
the GAMA limit in Fig.~\ref{fig:gama}. Use of the fainter, $K>18$,
VICS82 data to test LF parameters would increasingly depend on the
evolutionary model assumed. The bright limit is chosen because the
$Class~Star$ parameter is only calculated by \citet{Geach2017}  for
$K>12$ to avoid effects of saturation. The $K$ magnitudes are corrected 
into the 2MASS $K\_m\_ext$ system (see Section \ref{sub_subsec:vics82}
and Appendix \ref{appendix_vics82}). As also described in Section
\ref{sub_subsec:vics82} we have assumed a conservative star-galaxy
separation using $Class~Star<0.9$ and then removing any remaining
pointlike objects using Gaia data and eq \ref{eq:gaia}. We note that
there is good agreement with the counts given by \citet{Geach2017} in
their Fig. 5, once our magnitude offsets are taken into account. In the
full range, $12<K<16$, we again see excellent agreement with the
\citetalias{ws14} model and again the \citetalias{lh11} model
significantly under-predicts the galaxy counts. We conclude that, like
the GAMA counts, the VICS82 $K$-band data also strongly support the
accuracy of the \citetalias{ws14} model and its LF parameters, from
counts based on a completely independent sky area.

\subsection{Discussion}

What we observe is that the brighter $K<11.5$ 2MRS $n(z)$ requires a
20\% lower $\phi^*$ than the $K<15$ GAMA $n(z)$. So good fits to both
$n(z)$'s can be obtained if the LF $\phi^*$ is left as a free parameter
(see also Fig. 7 of \citealt{Sedgwick21}). This means that  the Local
Hole may have quite  a sharp spatial edge at $z\approx0.08$ or
$r\approx240h^{-1}$ Mpc. Otherwise, in an evolutionary interpretation
this would look more like pure density evolution than luminosity
evolution. In the density evolution case it is true that it would be
nearly impossible to differentiate a physical under-density from a
smoothly increasing galaxy density with redshift due to evolution. But
the reasonable fit of homogeneous models in the $z<0.08$ range would again
imply that there was a sharp jump in the galaxy density above this
redshift. Again this increase in density cannot continue at $z>0.08$ for
the same reason as for pure luminosity evolution, since the counts at
higher redshift would quickly be over-predicted. We regard either of
these sharply changing evolutionary scenarios around $z\approx0.08$ as
much less likely than an under-density, as has been argued for some
years even on the basis of blue-band number counts \citep{shanks90,Metcalfe91}.

We highlight the relative normalisations of the \citetalias{ws14} and
\citetalias{lh11} LF models as the key outcome of our analysis. The
\citetalias{lh11} model fails to fit the faint $n(m)$ galaxy counts in
the GAMA survey.  If their normalisation is correct and no local
under-density exists then it is implied that galaxies must evolve in a
way that their space density sharply increases at $z\ga0.08$ and $K>12$
and then returns to a non-evolving form out to $z\approx0.5$ and $K>20$.
This single spurt of evolution at $z\approx0.08$ has to be seen at
similar levels in the $b_J$, $r$ and $H$ bands as well as in the $K$
band. It was the unnaturalness of this evolutionary interpretation that
originally led e.g. \citet{shanks90} to normalise their LF estimates at
$b_J(\sim g)>17$ mag rather than at brighter magnitudes where the form
of the LF was estimated. Even authors who originally suggested such an
evolutionary explanation (e.g. \citealt{Maddox1990}) have more recently
suggested that a large scale structure explanation was more plausible
(e.g. \citealt{Norberg2002}). Moreover, \citetalias{ws14}  have
presented dynamical evidence for a local outflow in their analysis of
the relation between $\bar{z}$ and $m$ and \citet{shanks19, shanks19b}
have shown that this outflow is consistent with the Local Hole
under-density proposed here. It will also  be interesting to see whether
future all-sky SNIa supernova surveys confirm this $\bar{z}:m$ outflow
evidence, based as it is on the assumption that the $K-$band luminosity
function is a reasonable standard candle.
 
We suggest that the crucial  issue for \citetalias{lh11} and
\citet{Sedgwick21} is that they have fitted their LF parameters and
particularly the LF normalisation in the volume dominated by the Local
Hole and thus calibrated out the under-density. Certainly their $n(K)$ and
$n(z)$ models clearly fail at magnitudes and redshifts just outside the
ranges where they have determined their LF parameters. These authors
would need to show powerful evidence  for the $z<0.1$ evolution spurt in
the favoured $\Lambda$CDM model before their rejection of the Local Hole
hypothesis could be accepted. In the absence of such a model the balance
of evidence will clearly favour the Local Hole hypothesis.

\section{Conclusions}

In this work we have examined the local galaxy distribution and extended
the work of \citet{ws14} by measuring observed number-redshift $n(z)$
and number-magnitude $n(m)$ galaxy counts in the $K-$band across
$\approx90$\% of the sky down to a Galactic latitude
$|b|\ga5^\circ$. 

The $n(z)$ distributions from the 2MRS and 2M++ surveys to $K<11.5$ were
compared  to the homogeneous model of \citetalias{ws14} (see also
\citealt{metcalfe01,metcalfe06}). These wide-sky $n(z)$ distributions
showed excellent agreement and implied an under-density of $22\pm2$\%
relative to the model at $z<0.075$. We also find that the 2MASS $K$ counts
show a similar under-density of $20\pm2\%$ at $K<11.5$ relative to the
same model, only converging to the predicted count at $K\approx13.5$. In
addition, an LSS-corrected $\phi^{*}(z)$ model based on the $n(z)$
distribution, when compared to the 2MASS $K$ counts, showed a much
improved fit, confirming the consistency of the 2MASS $n(m)$ and the
2MRS/2M++ $n(z)$ in detecting this under-density relative to the
\citetalias{ws14} model. We also found the under-density covered 20/24
or $\approx83$\% of the observable wide-sky with only  areas containing
the Shapley and other  super-clusters and rich clusters like Coma showing up as
over- rather than an under-densities. 

Combined, our $n(m)$ and $n(z)$ counts are in good agreement with the
work of \citetalias{ws14}, \citet{frith03}, \citet{busswell04} and
\citet{keenan13}, who find overall underdensities of the order $\approx
15-25$ \% using  a similar galaxy counts method.  We also recall that in
the $\approx9000$deg$^2$ sky area analysed by \citetalias{ws14}, the
under-density patterns found in redshift were confirmed in detail by the
distribution traced by X-ray galaxy clusters  in the same volume
\citep{bohringer20}.

To examine whether our measured under-density represents a physical
Local Hole in the galaxy environment around our observer location
requires a confirmation of the accuracy of the \citetalias{ws14} galaxy
count model. We have investigated this by comparing the model's
predictions for the fainter  $K$ galaxy counts from the GAMA and VICS82
surveys. We have also compared these data with the model predictions of
\citetalias{lh11} who failed to find an under-density in the 2M++ survey. 

The $n(m)$ and $n(z)$ counts predicted by the \citetalias{lh11} model
are lower  by $\approx40$\% compared to the \citetalias{ws14} model; the
\citetalias{lh11} model thus initially appears to under-predict the 
observed wide-sky $n(K)$ and $n(z)$ distributions from 2MASS, 2MRS and
2M++. Then, at $K>13.5$, beyond the 2MASS sample range, the
\citetalias{ws14} prediction fits very well the  observed $n(K)$ and
$n(z)$ counts in the GAMA survey and the observed $n(K)$ in the VICS82
survey. However, the \citetalias{lh11} model shows a consistently poor
fit over both the full GAMA+VICS82	 $n(K)$ and GAMA $n(z)$
distributions. Thus the GAMA + VICS82 results indicate that the
\citetalias{ws14} model can more accurately fit deep $K-$ counts than
the \citetalias{lh11} model, supporting its use in interpreting the
lower redshift, wide-sky surveys.

Consequently, our analyses here support the existence of the `Local
Hole' under-density over $\approx90$\% of the sky. At the limiting
magnitude $K<11.5$ the under-density of $20\pm2$\% in the $n(z)$ counts
corresponds to a depth of $\approx 100h^{-1}$ Mpc, while the $13\pm1$\%
under-density at $K<12.5$ in the 2MASS wide-sky $n(m)$ counts, that is
in good agreement with \citetalias{ws14}, would imply the under-density
extends further to a depth of $\approx 150h^{-1}$Mpc. We note that
the statistical error on our LF normalisation  can be easily estimated
from the field-to-field errors in the $10<K<15$ galaxy counts between
the 3 GAMA fields (see Table \ref{tab:2MASS_GAMA}) and this gives an
error of $\pm3.4$\%. The error estimated from the two VICS82 sub-fields
would be similar at $\pm3.6$\% in the range $12<K<16$, decreasing to
$\pm1.1$\% in the range $12<K<18$.  Combining the GAMA $\pm3.4$\% error
with the $\pm2$\% error on the -20\% under-density to $K<11.5$ mag gives
the full uncertainty on the  Local Hole under-density out to 100h$^{-1}$
Mpc to be $-20\pm3.9$\% i.e. a $5.1\sigma$ detection. Similarly the
Local Hole $K<12.5$ under-density out to $\approx150$h$^{-1}$ Mpc is a
$-13\pm3.5$\% or a $3.7\sigma$ detection.

Such a 13-20\% underdensity at $\approx$100-150 h$^{-1}$Mpc scales would
notably affect distance scale measurements of the expansion rate $H_0$.
We can calculate this by assuming the linear theory discussed in
\citetalias{ws14} and \citet{shanks19}, where $\delta H_{0}/H_{0} =
-\frac{1}{3} \, \delta\rho_{g}/\rho_{g} \times \Omega_{m}^{0.6}/b$. Here
we take the galaxy bias $b\approx1.2$ for $K-$selected 2MRS galaxies in
the standard model (see e.g. \citealt{Boruah2020}; also
\citealt{Maller2005,Frith2005} although these latter $b$ values should
be treated as upper limits since they apply to $K<13.5$ and bias is
expected to rise with redshift.) From our measured $n(m)$ and $n(z)$
underdensities this would produce a decrease in the local value of $H_0$
of $\approx2-3$\%.

We finally consider the significance of such a large scale
inhomogeneity within the standard cosmological model. 
\citet{frith06} created mock 2MASS catalogues from the
Hubble Volume simulation to determine theoretically allowed
fluctuations and found that a $1\sigma$ fluctuation
to $H=13$ ($K\approx12.5$) over 65\% of the sky corresponded to
$\pm3.25$\%. Scaling this to the 90\% wide-sky coverage used here
implies $1\sigma=2.8$\%. Given our $13\pm3.5$\% under-density to
$K<12.5$, we can add in quadrature this $\pm2.8$\%
expected fluctuation from the $\Lambda$CDM model to obtain $13\pm4.5$\%
with the error now including our measurement error and the expected
count fluctuation expected out to $\approx150$h$^{-1}$ in  $\Lambda$CDM.
The Local Hole with a 13\% under-density therefore here corresponds to
a $2.9\sigma$ deviation from what is expected in a  $\Lambda$CDM cosmology.

If we scale this from $K<12.5$ mag to $K<11.5$ mag via a 3-D version of 
Eq. 3 of \citet{frith05}, a $1\sigma$ fluctuation at $K<11.5$ corresponds
to $\pm5.6$\%. At $K<12.5$, the under-density is $-20\pm2$\% and  folding in
the $\pm3.4$\% normalisation error gives $-20\pm3.9$\% or a $5.1\sigma$
detection of the Local Hole under-density. Then adding in the $\pm5.6$\%
expected fluctuation amplitude just calculated gives $-20\pm6.8$\%,
implying again a $2.9\sigma$ deviation in the $\Lambda$CDM cosmology,
similar to the $K<11.5$ case.

However, the deviation from $\Lambda$CDM is likely to be more
significant. For example, if we normalised our model via  the VICS82
$n(K)$ counts in the $12<K<18$ range (see Fig. \ref{fig:vics82}) then
this would argue that our LF normalisation should be still higher and the
field-field error would also be lower at $\pm1.1$\%. Additionally,
taking into account the excellent fit of the WS14 model  to the 2MASS
wide-sky data itself at $12.5<K<13.5$ (see Fig. \ref{fig:nm}) would also
further increase the significance of the deviation from $\Lambda$CDM.

Although the  Hubble Volume mocks of \citet{frith06} have
tested our methodology in the context of an N-body simulation `snapshot'
with an appropriate galaxy clustering amplitude  in volumes similar to
those sampled here, it would be useful to make further tests in a more
realistic simulation. For example, a full lightcone analysis could be
made, applying our  selection cuts in a mock that includes a full
`semi-analytic' galaxy formation  model (e.g. \citealt{Sawala2022}).
This would make a further direct test of our methodology while checking
if there is any evolutionary effect that provides the spurt of density
evolution at $z\approx0.08$ required to provide an alternative to our
large-scale clustering explanation of the Local Hole.

We therefore anticipate that further work to separate out the effects of
evolution and LSS on the luminosity function in each of the
\citetalias{ws14} and \citetalias{lh11} approaches will shed further
light on the presence and extent of the Local Hole. Similarly, further
work will be needed to resolve the discrepancy between the detection of
dynamical infall at the appropriate level implied from the Local Hole
under-density found by \citetalias{ws14}, \citet{shanks19} and
\citet{shanks19b} as compared to the lack of such infall found by
\citet{kenworthy19} and \citet{Sedgwick21}. But here we have confirmed
that the proposed Local Hole under-density extends to cover almost the
whole sky, and argued that previous failures to find the under-density
are generally due to homogeneous number count models that assume global
LF normalisations that are biased low by being determined within the
Local Hole region itself. 

Finally, if the form of the galaxy $n(K)$ and $n(z)$ do
imply a `Local Hole' then how could it fit into the standard
$\Lambda$CDM cosmology? Other authors have suggested possibilities to
explain unexpectedly large scale inhomogeneities such as an anisotropic
Universe (e.g. \citealt{Secrest2021}). However, it is hard to see how such
suggestions retain the successes of the standard model in terms of the
CMB power spectrum etc. We note that other anomalies in the local galaxy
distribution exist e.g. \citet{Mackenzie2017} presented evidence for a
coherence in the galaxy redshift distribution across
$\approx600$h$^{-1}$ Mpc of the Southern sky out to $z\approx0.1$.
Prompted by this result and by the `Local Hole' result reported here,
Callow et al. (2021, in prep.) will discuss the possibilities that arise
if the topology of the Universe is not simply connected. We emphasise
that there is no proof but here we just use this model as an example of one
that might retain the basic features of the standard model while
producing a larger than expected coherent local under- or over-density.
It will be interesting to look for other models that introduce such `new
physics` to explain the local large-scale structure while simultaneously
reducing the tension in Hubble's Constant.

\section*{Acknowledgements}

We first acknowledge the comments of an anonymous referee that have significantly
improved the quality of this paper. We further acknowledge STFC Consolidated Grant 
ST/T000244/1 in supporting this research.

This publication makes use of data products from the Two Micron All Sky
Survey (2MASS), which is a joint project of the University of
Massachusetts and the Infrared Processing and Analysis Center/California
Institute of Technology, funded by the National Aeronautics and Space
Administration and the National Science Foundation.

It also makes use of the 2MASS Redshift survey catalogue as described by
\citet{huchra12}. The version used here is catalog version 2.4 from the
website http://tdc-www.harvard.edu/2mrs/ maintained by Lucas Macri.

Funding for SDSS-III has been provided by the Alfred P. Sloan
Foundation, the Participating Institutions, the National Science
Foundation and the US Department of Energy Office of Science. The
SDSS-III website is http://www.sdss3.org/.

The 6dF Galaxy Survey is supported by Australian Research Council
Discovery Projects Grant (DP-0208876). The 6dFGS web site is
http://www.aao.gov.au/local/www/6df/.

GAMA is a joint European–Australasian project based around a
spectroscopic campaign using the Anglo-Australian Telescope. The GAMA
input catalogue is based on data taken from the Sloan Digital Sky Survey
and the UKIRT Infrared Deep Sky Survey. Com- plementary imaging of the
GAMA regions is being obtained by a number of independent survey
programmes including GALEX MIS, VST KiDS, VISTA VIKING, WISE,
Herschel-ATLAS, GMRT and ASKAP providing UV to radio coverage. GAMA is
funded by the STFC (UK), the ARC (Australia), the AAO and the
participating institutions. The GAMA website is
http://www.gama-survey.org/.

This work has made use of data from the European Space Agency (ESA) mission
{\it Gaia} (\url{https://www.cosmos.esa.int/gaia}), processed by the {\it Gaia}
Data Processing and Analysis Consortium (DPAC,
\url{https://www.cosmos.esa.int/web/gaia/dpac/consortium}). Funding for the DPAC
has been provided by national institutions, in particular the institutions
participating in the {\it Gaia} Multilateral Agreement.

\section*{Data Availability}
The 2MASS, 2MRS, 6dF, SDSS, GAMA, VICS82 and Gaia data we have used are all publicly available.
All other data relevant to this publication will be supplied on request to the authors. 



\bibliographystyle{mnras}
\bibliography{ref} 




\appendix

\section{GAMA-2MASS magnitude comparison}
\label{appendix_gama}

\begin{figure}
\centering
\caption{Differences between 2MASS $K\_m\_ext$ and  GAMA
\textit{MAG\_AUTO\_K} as a function of GAMA magnitudes for each GAMA field.}
\includegraphics[width=\columnwidth]{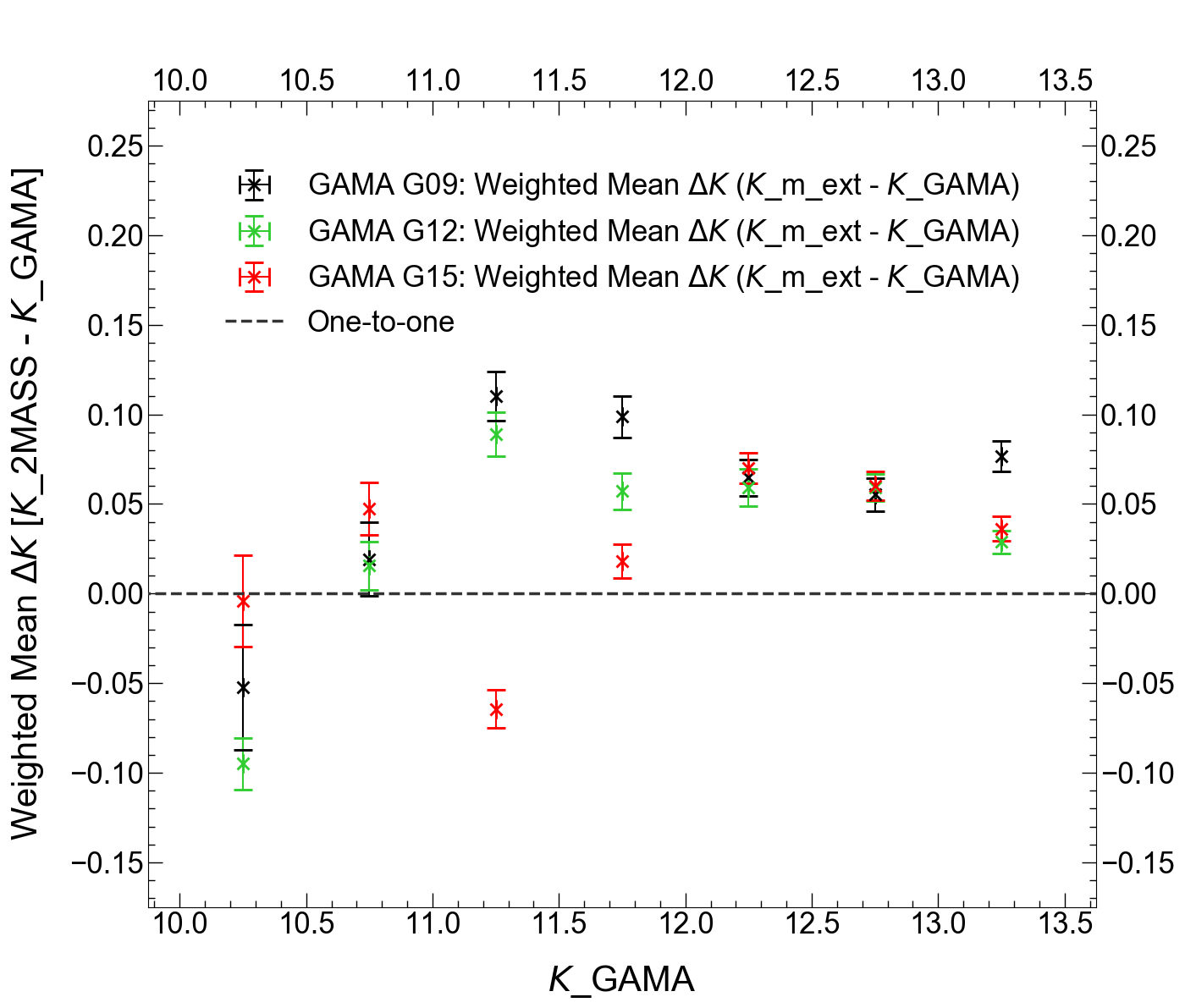}
\label{fig:ext_gama}
\end{figure}

\begin{figure}
\centering
\caption{Differences between 2MASS $K\_BEST$ and  GAMA
\textit{MAG\_AUTO\_K} as a function of GAMA magnitudes}
\includegraphics[width=\columnwidth]{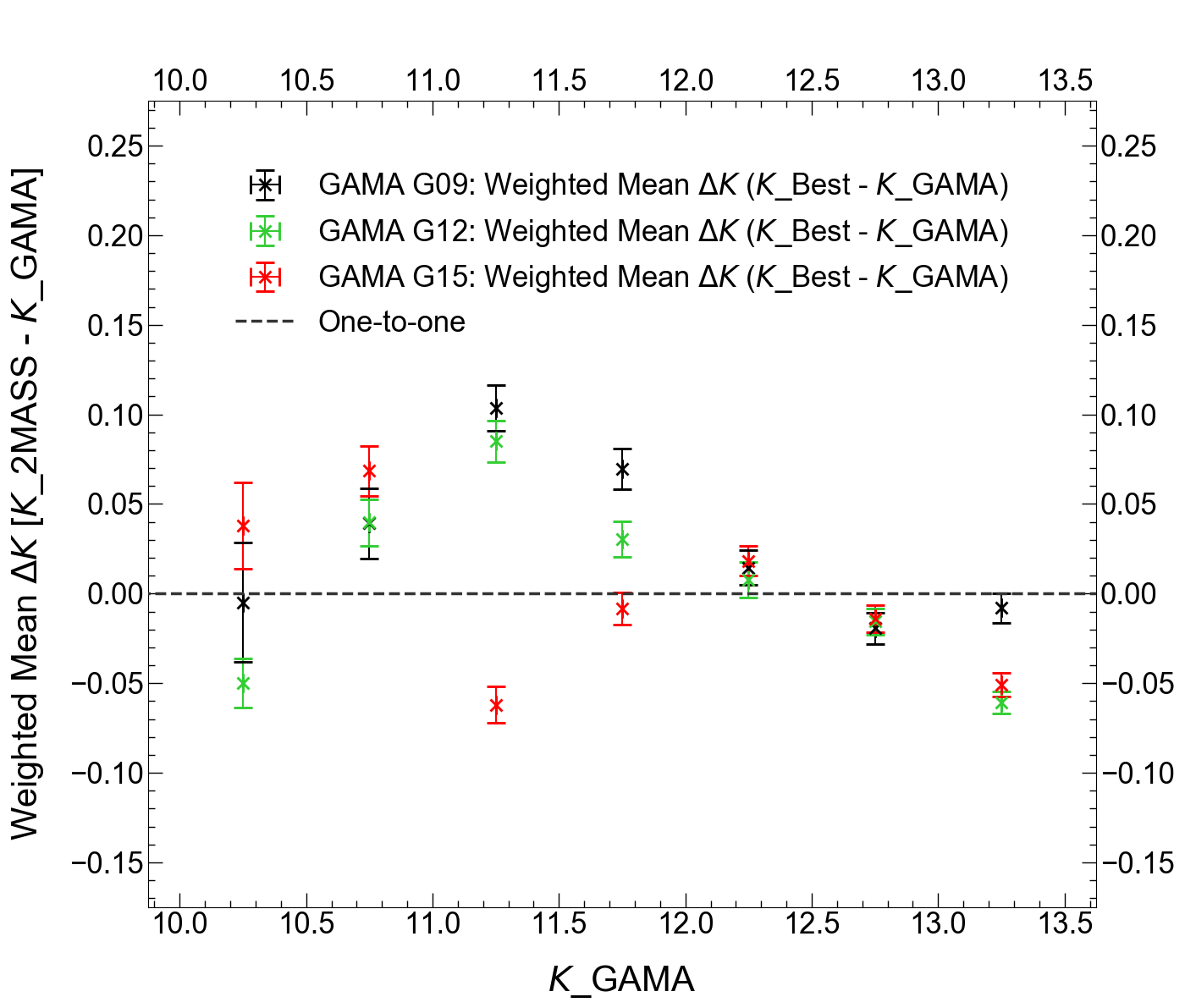}
\label{fig:best_gama}
\end{figure}

We first show Fig. \ref{fig:ext_gama} where GAMA
\textit{MAG\_AUTO\_K} and 2MASS $K\_m\_ext$ magnitudes are  directly
compared. Table \ref{tab:2MASS_GAMA} shows the error weighted mean of the differences between
these two for each GAMA field within the range $10<K\_GAMA<13.5$ mag.
Similarly Fig. \ref{fig:best_gama} shows the comparison between GAMA
\textit{MAG\_AUTO\_K} and 2MASS magnitudes and  Table \ref{tab:2MASS_GAMA} 
again shows the error weighted mean of the differences for each field in the same $K\_GAMA$ range 

Following WS14, we have conservatively corrected the GAMA magnitudes for each of the
three fields by correcting  the GAMA magnitudes by adding  the 2MASS $K\_m\_ext$
magnitude offsets given in the fourth column of Table \ref{tab:2MASS_GAMA} rather than the 
2MASS $k\_BEST$ offsets given in the third column. This takes the GAMA magnitudes into the 2MASS
$K\_m\_ext$ system rather than the $K\_BEST$ system we are actually using. Clearly if we used the 
$K\_BEST$ offsets the GAMA $K$ counts would lie even higher in Fig. \ref{fig:gama}.

\begin{table}
\begin{center}
\caption{The magnitude offsets needed to correct the GAMA
\textit{MAG\_AUTO\_K} ($=K\_GAMA$) in each of the GAMA fields into the
2MASS $K\_BEST$  and $K\_m\_ext$ systems. They were calculated by taking
an error weighted mean of the magnitude differences in $10<K_{GAMA}<13.5$ mag.}
\label{tab:2MASS_GAMA}
\begin{tabular}{lccc}
\hline
\hline
GAMA       & N (2MASS              & Weighted Mean          & Weighted Mean\\
Field      & $\times$GAMA)         & $K_{Best} - K_{GAMA}$  & $K\_m\_ext - K_{GAMA}$\\
\hline
\hline
G09 & 876   & $0.019\pm0.004$     &   $0.071\pm0.004$\\
G12 & 1,208 & $-0.011\pm0.003$    &   $0.039\pm0.004$\\
G15 & 1,184 & $-0.017\pm0.004$    &   $0.033\pm0.004$\\
\hline
\hline
\end{tabular}
\end{center}
\end{table}

\section{2MASS magnitude comparison}
\label{appendix_2mass}

\begin{figure}
\centering
\caption{The 2MASS magnitude offsets from Table \ref{tab:ext_k20} between the $K\_m\_ext$ magnitudes on which WS14 and our 
magnitudes are based and the $K\_20$ magnitudes used by \citetalias{lh11}.}
\includegraphics[width=\columnwidth]{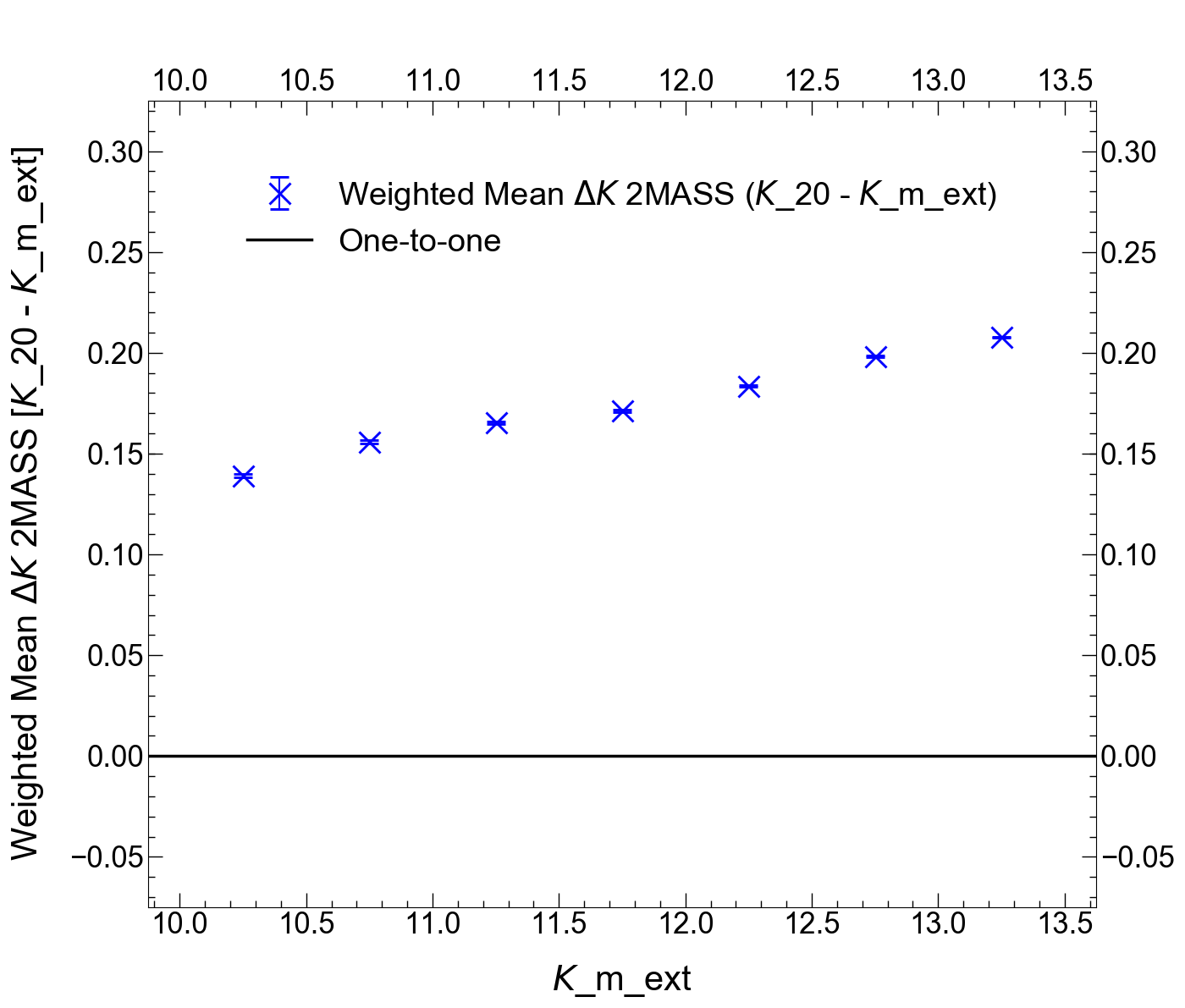}
\label{fig:ext_k20}
\end{figure}

Here we compare the 2MASS magnitude system, $K\_m\_ext$, on which our
and \citetalias{ws14} $n(K)$ results are based with the 2MASS
$K\_20(=K\_Kron)$ magnitudes used by \citetalias{lh11}. The comparisons
are shown as a function of $K\_m\_ext$ in Table \ref{tab:ext_k20} and
Fig. \ref{fig:ext_k20}. The overall difference is found to be
$K\_m\_ext - K\_20=0.19\pm 0.0002$ mag.

\begin{table}
\begin{center}
\caption{The 2MASS magnitude offsets between the $K\_m\_ext$ magnitudes on which WS14 and our 
magnitudes are based and the $K\_20$ magnitudes used by \citetalias{lh11} (see Fig. \ref{fig:ext_k20}).}
\label{tab:ext_k20}
\begin{tabular}{ccc}
\hline
\hline
Magnitude Range          &$N_{gal}$& Weighted Mean\\
                         &2MASS    & $K\_m\_ext - K\_20$\\
\hline
\hline
$10.0<K\_m\_ext<10.5$ & 4,859   &   $0.139\pm0.0008$\\
$10.5<K\_m\_ext<11.0$ & 9,827   &   $0.156\pm0.0007$\\
$11.0<K\_m\_ext<11.5$ & 18,846  &   $0.166\pm0.0006$\\
$11.5<K\_m\_ext<12.0$ & 38,756  &   $0.171\pm0.0005$\\
$12.0<K\_m\_ext<12.5$ & 81,227  &   $0.184\pm0.0004$\\
$12.5<K\_m\_ext<13.0$ & 169,857 &   $0.198\pm0.0004$\\
$13.0<K\_m\_ext<13.5$ & 359,686 &   $0.208\pm0.0003$\\
\hline
$10.0<K\_m\_ext<13.5$ & 683,958 &   $0.188\pm0.0002$\\
\hline
\hline
\end{tabular}
\end{center}
\end{table}

\section{2MASS-VICS82 magnitude comparison}
\label{appendix_vics82}

\begin{figure}
\centering
\caption{The offsets between the 2MASS $K\_m\_ext$ magnitudes on which \citetalias{ws14} and our 
magnitudes are based and the VICS82 $MAG\_AUTO$ magnitudes used by \citet{Geach2017} (see Table \ref{tab:ext_vics82}).}
\includegraphics[width=\columnwidth]{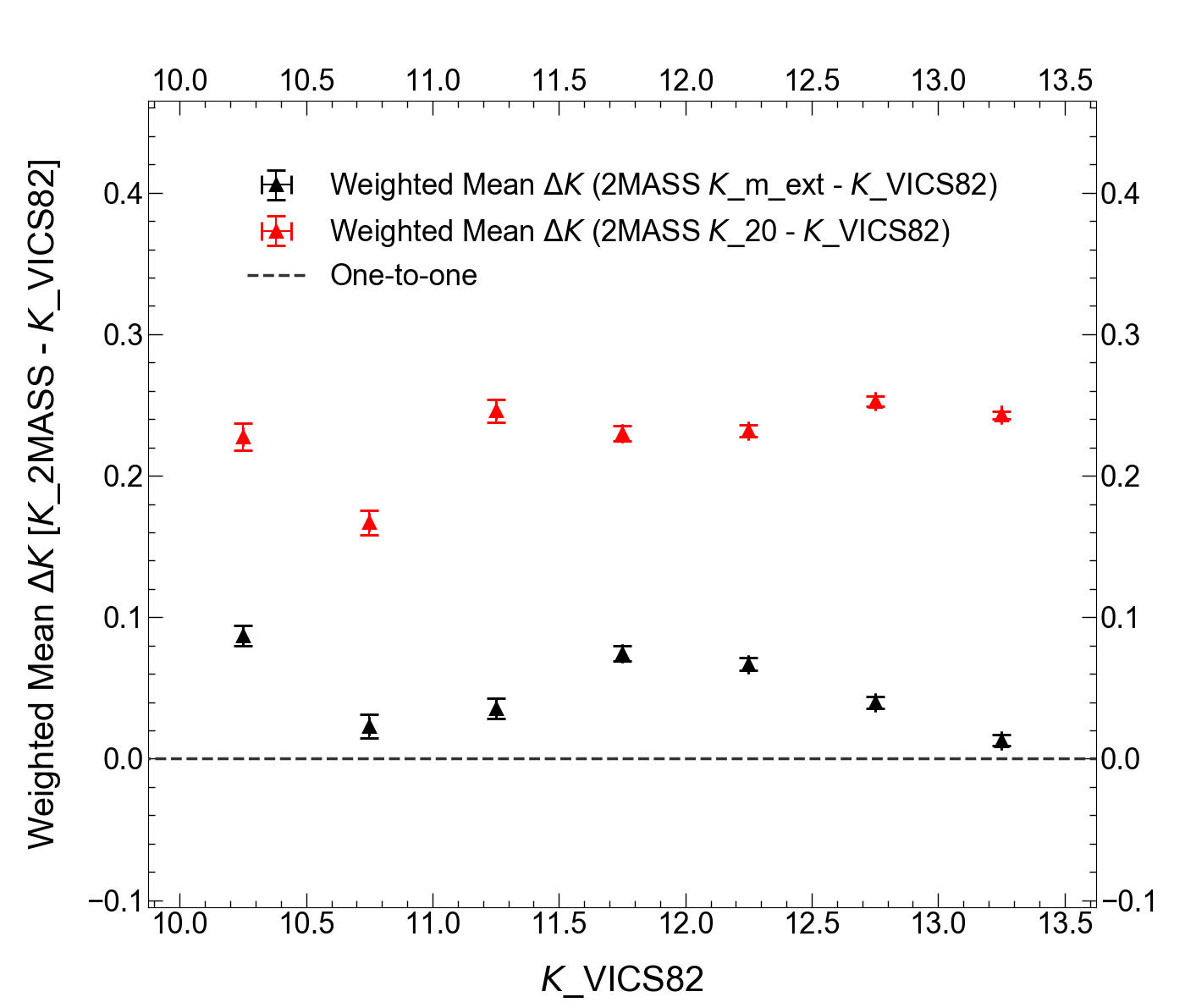}
\label{fig:ext_vics82}
\end{figure}

Table \ref{tab:ext_vics82} and Fig. \ref{fig:ext_vics82} shows the offsets between] 
the $K\_m\_ext$ and VICS82 $MAG\_AUTO$ magnitude systems. Because of the possibility of 
saturation affecting the VICS82 magnitudes at $K<12.5$ and poor S/N affecting the fainter 
2MASS magnitudes we simply average the 3 values in the range $12.5<K\_m\_ext<13.5$ to obtain the overall offset
$K\_m\_ext - K\_VICS82=0.040\pm 0.004$ mag as used in Section \ref{sub_subsec:vics82}.

\begin{table}
\begin{center}
\caption{The offsets between the 2MASS $K\_m\_ext$ magnitudes on which \citetalias{ws14} and our 
magnitudes are based and the VICS82 $MAG\_AUTO$ magnitudes used by \citet{Geach2017} (see Fig. \ref{fig:ext_vics82}).}
\label{tab:ext_vics82}
\begin{tabular}{ccc}
\hline
\hline
Magnitude Range          &$N_{gal}$& Weighted Mean\\
                         &2MASS$\times$VICS82   & $K\_m\_ext - K\_VICS82$\\
\hline
\hline
$10.0<K\_m\_ext<10.5$ & 26      &   $0.087\pm0.007$\\
$10.5<K\_m\_ext<11.0$ & 31      &   $0.023\pm0.009$\\
$11.0<K\_m\_ext<11.5$ & 73      &   $0.036\pm0.007$\\
$11.5<K\_m\_ext<12.0$ & 175     &   $0.075\pm0.005$\\
$12.0<K\_m\_ext<12.5$ & 361     &   $0.067\pm0.005$\\
$12.5<K\_m\_ext<13.0$ & 730     &   $0.040\pm0.004$\\
$13.0<K\_m\_ext<13.5$ & 1536    &   $0.013\pm0.004$\\
\hline
$10.0<K\_m\_ext<13.5$ & 2932    &   $0.044\pm0.002$\\
\hline
\hline
\end{tabular}
\end{center}
\end{table}

\section{Spectroscopic Incompleteness of $n(z)$ Counts}
\label{appendix_spectra}

\begin{figure}
\centering
\caption{The $K-$band spectroscopic completeness of 2MRS and 2M++ with
respect to the 2MASS Extended Source Catalogue, evaluated per half
magnitude bin to the limit $K<11.5$. Errors have been calculated using
the field-field method.}
\includegraphics[width=\columnwidth]{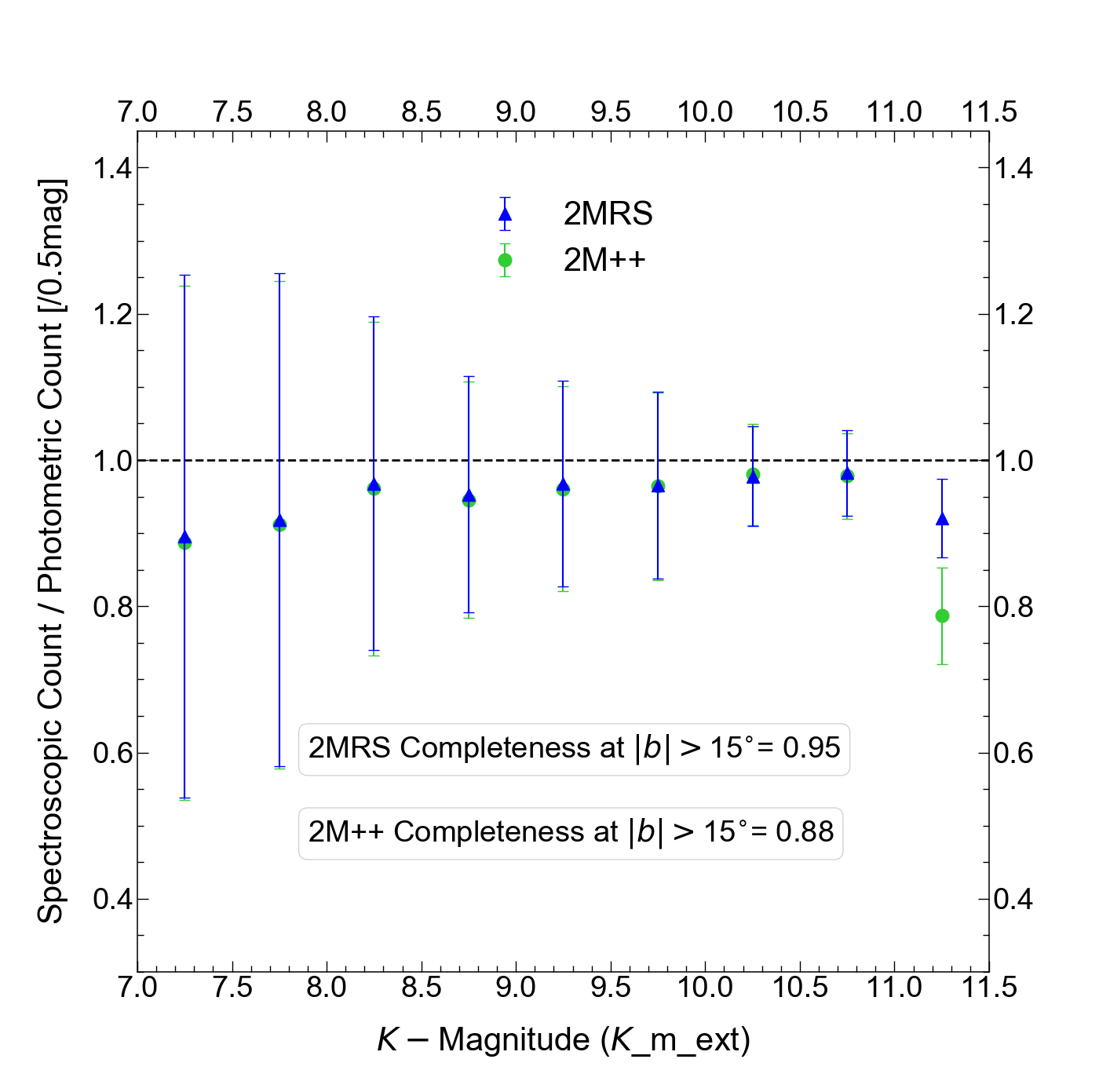}
\label{Figure:A1}
\end{figure}

In Fig. \ref{Figure:A1} we present the calculation of the spectroscopic
incompleteness factors applied to 2MRS and 2M++ $n(z)$ data before
fitting to the \citetalias{ws14} homogeneous model. The $n(z)$ samples
are matched to the 2MASS Extended Source Catalogue, and we plot the
ratio of the $n(z)$ galaxy count to the 2MASS galaxy count, per half
magnitude bin.

The overall completeness of each survey is calculated by the total sum
of spectroscopic sources in either 2MRS or 2M++, divided by the total
sum of photometric sources in 2MASS taken over the full magnitude range.
We measure a completeness of $95\%$ in 2MRS and $88\%$ in 2M++, and the
reciprocal of these values is the spectroscopic incompleteness factor
which is multiplied to the raw $n(z)$ data of each survey.

\bsp	
\label{lastpage}
\end{document}